\documentclass[twocolumn]{aastex631} 
\usepackage{newtxtext,amsfonts}

\usepackage[T1]{fontenc}

\DeclareRobustCommand{\VAN}[3]{#2}
\let\VANthebibliography\thebibliography
\def\thebibliography{\DeclareRobustCommand{\VAN}[3]{##3}\VANthebibliography}

\usepackage{graphicx}	
\usepackage{amsmath}	
\usepackage{float}
\usepackage[autopunct=true]{csquotes}
\usepackage{stfloats}
\usepackage{blindtext}
\usepackage{cuted}
\graphicspath{{./}{figures/}}

\begin{document}
\newcommand{\pfh}[1]{\textcolor{red}{#1}}

\newcommand{\GIZMO}{{\small GIZMO }}
\newcommand{\SF}{{\small STARFORGE }}

\newcommand{\gizmourl}{\href{http://www.tapir.caltech.edu/~phopkins/Site/GIZMO.html}{\url{http://www.tapir.caltech.edu/~phopkins/Site/GIZMO.html}}}

\newcommand{\datastatement}[1]{\begin{small}\section*{Data Availability Statement}\end{small}{\noindent #1}\vspace{5pt}}
\newcommand{\microGauss}{\mu{\rm G}}
\newcommand{\Bangle}{\theta_{B}}
\newcommand{\Alf}{{Alfv\'en}}
\newcommand{\BV}{Brunt-V\"ais\"al\"a}
\newcommand{\fref}[1]{Fig.~\ref{#1}}
\newcommand{\sref}[1]{\S~\ref{#1}}
\newcommand{\aref}[1]{App.~\ref{#1}}
\newcommand{\tref}[1]{Table~\ref{#1}}

\newcommand{\Dt}[1]{\frac{\mathrm{d} #1}{\mathrm{dt}}}
\newcommand{\initvalupper}[1]{#1^{0}}
\newcommand{\initvallower}[1]{#1_{0}}
\newcommand{\driftvel}{{\bf w}_{s}}
\newcommand{\driftvelmag}{w_{s}}
\newcommand{\driftvelhat}{\hat{{\bf w}}_{s}}
\newcommand{\driftveli}[1]{{\bf w}_{s,\,#1}}
\newcommand{\driftvelmagi}[1]{w_{s,\,#1}}
\newcommand{\dustvel}{{\bf v}_{d}}
\newcommand{\gasvel}{{\bf u}_{g}}
\newcommand{\gasden}{\rho_{g}}
\newcommand{\rhobase}{\rho_{\rm base}}
\newcommand{\gaspressure}{P}
\newcommand{\dustden}{\rho_{d}}
\newcommand{\rhodust}{\dustden}
\newcommand{\rhogas}{\gasden}
\newcommand{\resolution}{\Delta x_{0}}
\newcommand{\opticaldepth}{\tau_{\rm ext}}
\newcommand{\tauparam}{\tau_{\rm SL}}
\newcommand{\ts}{t_{s}}
\newcommand{\cs}{c_{s}}
\newcommand{\vA}{v_{A}}
\newcommand{\tL}{t_{L}}
\newcommand{\grainsuff}{_{\rm grain}}
\newcommand{\internaldensity}{\bar{\rho}\grainsuff^{\,i}}
\newcommand{\grainsize}{\epsilon\grainsuff}
\newcommand{\grainsizebar}{\bar{\epsilon}\grainsuff}
\newcommand{\grainmass}{m\grainsuff}
\newcommand{\graincharge}{q\grainsuff}
\newcommand{\grainchargeZ}{Z\grainsuff}
\newcommand{\grainsizemax}{\grainsize^{\rm max}}
\newcommand{\grainsizemin}{\grainsize^{\rm min}}
\newcommand{\B}{{\bf B}}
\newcommand{\Bmag}{|\B|}
\newcommand{\Bhat}{\hat\B}
\newcommand{\bhat}{\Bhat}
\newcommand{\acc}{{\bf a}}
\newcommand{\Lbox}{L_{\rm box}}
\newcommand{\Lscale}{H_{\rm gas}}
\newcommand{\sizeparam}{\tilde{\alpha}}
\newcommand{\sizeparammax}{\sizeparam_{\rm m}}
\newcommand{\chargeparam}{\tilde{\phi}}
\newcommand{\chargeparammax}{\chargeparam_{\rm m}}
\newcommand{\accparam}{\tilde{a}_{\rm d}}
\newcommand{\accparammax}{\tilde{a}_{\rm d,m}}
\newcommand{\accabsmax}{{a}_{\rm d,m}}
\newcommand{\accsizedep}{\psi_{a}}
\newcommand{\gravparam}{\tilde{g}}
\newcommand{\dustgas}{\mu^{\rm dg}}
\newcommand{\dustgashat}{\hat{\mu}^{\rm dg}}
\newcommand{\angstrom}{\mbox{\normalfont\AA}}

\def\lesssimA#1#2{\mathrel{\vcenter{\offinterlineskip%
    \ialign{\hfil##\hfil\cr$#1<$\cr$#1\sim$\cr}%
}}}

\def\lesssim{\mathpalette\lesssimA{}}

\def\app#1#2{%
  \mathrel{%
    \setbox0=\hbox{$#1\sim$}%
    \setbox2=\hbox{%
      \rlap{\hbox{$#1\propto$}}%
      \lower1.1\ht0\box0%
    }%
    \raise0.25\ht2\box2%
  }%
}
\def\approxprop{\mathpalette\app\relax}

\title{Dust Battery: A Novel Mechanism for Seed Magnetic Field Generation in the Early Universe}
\shorttitle{Dust Battery}
\correspondingauthor{Nadine H.~Soliman}
\email{nsoliman@caltech.edu}
\author[0000-0002-6810-1110]{Nadine H.~Soliman}
\affiliation{TAPIR, Mailcode 350-17, California Institute of Technology, Pasadena, CA 91125, USA}
\author[0000-0003-3729-1684]{Philip F.~Hopkins}
\affiliation{TAPIR, Mailcode 350-17, California Institute of Technology, Pasadena, CA 91125, USA}
\author[0000-0001-8479-962X]{Jonathan Squire}
\affiliation{Physics Department, University of Otago, 730 Cumberland St., Dunedin 9016, New Zealand}

\shortauthors{Soliman et al.}

\begin{abstract}
We propose a novel dust battery mechanism for generating seed magnetic fields in the early universe, in which charged dust grains are radiatively accelerated, inducing strong electric currents that subsequently generate magnetic fields. Our analysis demonstrates that this process is effective even at very low metallicities (approximately $ \sim 10^{-5} Z_\odot$), and capable of producing seed fields with significant amplitudes of $B \sim \rm \mu G$ around luminous sources over timescales of years to Myr and across spatial scales ranging from AU to kpc. Crucially, we find that this mechanism is generically $\sim10^8$ times more effective than the radiatively-driven electron battery or Biermann battery in relatively cool gas ($\ll 10^{5}\,$K), including both neutral and ionized gas. Furthermore, our results suggest that, to first order, dissipation effects do not appear to significantly impede this process, and that it can feasibly generate coherent seed fields on macroscopically large ISM scales (much larger than turbulent dissipation scales or electron mean-free-paths in the ISM). These seed fields could then be amplified by subsequent dynamo actions to the observed magnetic fields in galaxies. Additionally, we propose a sub-grid model for integration into cosmological simulations, and the required electric-field expressions for magnetohydrodynamic-particle-in-a-cell (MHD-PIC) simulations that explicitly model dust dynamics. Finally, we explore the broad applicability of this mechanism across different scales and conditions, emphasizing its robustness compared to other known battery mechanisms. 
\end{abstract}

\keywords{Cosmology (343); Galaxy formation (595); Magnetic fields (994); Magnetohydrodynamics (1964); Cosmic magnetic fields theory (321); Astrophysical dust processes (99); Computational methods (1965)}

\section{Introduction} \label{sec:intro}

Magnetic fields with strengths in the nanogauss to milligauss range have been observed across various scales and structures both in the present-day universe and at higher redshifts  \citep{kronberg1994extragalactic, athreya1998large, widrow2002origin, carilli2002cluster, kulsrud_zweibel, bernet2008strong, beck2011cosmic, vallee2011magnetic, ryu2012magnetic, feretti2012clusters, beck2013magnetic, ferrario2015magnetic, beck2016new}. However, the origin of magnetic fields in the universe remains a longstanding mystery in astrophysics. Dynamos are capable of amplifying extremely weak seed fields (as low as $\sim 10^{-23}-10^{-19} \rm G$) to observed strengths and extending these fields from localized sources to the intergalactic medium \citep{pudritz1989origin, tan2004protostellar, silk2006first, schleicher2010small, sur2010generation}. Turbulent and small-scale dynamos enhance magnetic fields within dense regions such as the interstellar medium of galaxies \citep{dolag1999sph, arshakian2009evolution, federrath2011mach, beresnyak2016turbulent, federrath2016magnetic}, while galactic-scale dynamos, driven by processes like fountains and winds \citep{hanasz2004building, rieder2016small, rieder2017small, pakmor2017magnetic}, contribute to large-scale amplification. Although the efficiency of these dynamo mechanisms varies, they all fundamentally rely on the presence of an initial non-zero seed field generated by a plasma physics process.

Theories for the generation of these seed fields generally fall into two main categories: cosmogenic fields and late-Universe battery mechanisms. Cosmogenic fields are hypothesized to arise from early universe phenomena during inflation \citep{ratra1992cosmological, turner1988inflation, campanelli2013origin} from Grand Unified Theory (GUT) scale physics \citep{ quashnock1989magnetic, grasso2001magnetic, vachaspati2008magnetic, kandus2011primordial,durrer2013cosmological}. While these models offer interesting possibilities, they typically invoke physics beyond the Standard Model introducing significant theoretical uncertainties. Direct observational support for such models remains limited. However, lower limits on intergalactic magnetic fields, as reported by \citet{neronov2010evidence}, challenge conventional astrophysical explanations, as they cannot be readily explained by known mechanisms. Recent work by \citet{tjemsland2024constraining} further demonstrates that standard scenarios struggle to reproduce these observed limits, underscoring the potential role of cosmogenic processes and the need for new physics or alternative theoretical frameworks.

Alternatively, several battery mechanisms have been proposed to generate seed magnetic fields in the late Universe by inducing some charge separation. Each has their own strengths and limitations. Among the most well-known are (i) the Biermann battery \citep{biermann}, which generates fields through misaligned gradients of electron pressure and number density; (ii) kinetic instabilities like the Weibel instability \citep{weibel1959spontaneously, califano1997spatial, medvedev1999generation}, which produce and amplify fields from anisotropies in particle velocity distributions; and (iii) radiation-driven batteries \citep{harrison1973magnetic, langer2004large, ando2010generation, durrive2015intergalactic}, which rely on electron opacity to background radiation. However, these mechanisms typically produce relatively weak seed magnetic fields on small scales and on specific conditions, such as highly ionized and high-temperature environments for the Wiebel instability, that were not prevalent in the early Universe \citep{widrow2002origin}. 

This raises significant questions about whether these mechanisms alone can account for the widespread and strong magnetic fields observed across the Universe. Observations of high-redshift galaxies ($z \sim 2.6$) reveal magnetic field strengths as high as $\lesssim 500 \, \rm \mu G$ \citep{geach2023polarized}, which would require efficient dynamo amplification if originating from weak seed fields. While dynamo processes could plausibly amplify these fields over cosmic timescales, they could also arise due to stronger initial seeding mechanisms capable of producing substantial magnetic fields at earlier epochs.

Cosmological simulations have also explored the possibility of magnetic field seeding by stellar or supernova sources \citep{star, pudritz1989origin} and active galactic nuclei (AGN) \citep{daly1990possible, furlanetto2001intergalactic}. However, these models still implicitly depend on unresolved battery mechanisms, such as those described earlier, and then simply assume some efficient battery in these environments. 

In this context, we propose a new mechanism for generating seed magnetic fields in the early universe through a dust battery process involving charged dust grains. We demonstrate that this mechanism can be highly efficient and capable of operating in environments where other mechanisms may fail. Observational evidence strongly supports the presence of significant amounts of dust in the high-redshift universe, with detections in galaxies at redshifts as high as  $z \sim 8$ \citep{laporte2017dust, tamura2019detection, viero2022early, inami2022alma, dayal2022alma, witstok2023empirical}. These observations underscore the abundance of dust during this epoch, highlighting its essential role in early galaxy evolution and positioning it as a plausible contributor to the generation of seed magnetic fields.  

This paper is organized as follows: In \S \ref{sec:form}, we present the governing equations the multifluid dynamics, and derive the electric field induced by radiation pressure accelerating charged dust grains. \S \ref{sec:tva} applies this formalism to specific astrophysical environments, where we compute the resultant seed magnetic field and identify the conditions under which it can be efficiently generated. In \S \ref{sec:subgrid}, we present a sub-grid model aimed at capturing the unresolved small-scale physics of the battery, which can be integrated into large-scale cosmological simulations. \S {\ref{sec:comp} presents a comparative analysis of this mechanism with other known battery processes. Finally, in \S \ref{sec:conc}, we summarize the key results and implications of our study.

\section{Formalism}
\label{sec:form}
We study the generation of magnetic fields from a current generated by radiative acceleration on charged dust grains in a mostly neutral medium.

\subsection{Individual continuity and momentum equations }
\label{sec:general.equations}

We begin with the multifluid continuum equations for an arbitrary number of charged dust and gas species as in \citet{cowling:1976.mhd.book, ichimaru:1978.nonideal.mhd.deriv.and.assumptions, nakano:1986.nonideal.mhd.formulation.review}. We first derive the governing equations for the ``\textit{gas}'' components (all non-dust species) and express the induced electric field that drives the generation of a magnetic field, in terms of an arbitrary dust current. In \S \ref{sec:tva}, we then introduce additional assumptions to approximate the dust current and compute the resulting magnetic field seeding rates. The most general solutions are not generally instructive (involving non-closed-form expressions), but we find all of the key behaviors and dimensional scalings of interest are captured by reducing to a four-plus-component system of free electrons ($e$), positively charged ions ($+$), neutrals ($n$), and charged dust\footnote{Throughout, quantities like $n_{d}$ implicitly refer to the number of {\em charged} dust grains. In well-ionized environments, this will be essentially all dust grains \citep{umebayashi1980recombination,nakano:1986.nonideal.mhd.formulation.review, nakano2002mechanism}, but in highly neutral environments there is sufficiently little free charge that some grains will be uncharged even if grains absorb all the free charge. One could represent neutral grains by summing over grain sub-populations with different charge $q_{d}$ (including $q_{d}=0$) in our expressions, but to leading order they have no effect.} ($d$, which can represent a sum over many dust sizes/species), to which we reduce below. Each species $j$ is characterized by its particle mass $m_j$, number density $n_j$, mass density $\rho_j = n_j m_j$, and signed charge $q_j = \pm Z_j q_e$, where $q_e$ represents the elementary charge and $Z_j$ is the charge number. The microscopic velocity of each species is denoted by ${\bf{v}}_j$, and the particles experience an external acceleration ${\bf a}_{{\rm ext},j}$. The mean velocity, averaged over the distribution function, is given by ${\bf u}_j \equiv \langle {\bf{v}}_j \rangle$. Here, $\langle {\bf{x}}_j \rangle \equiv \int {\bf{x}}_j f_j {\rm d}{\bf{x}}_j$ denotes the average over the distribution function $f_j$, normalized such that $\int f_j {\rm d}{\bf{x}}_j = 1$.

We consider infinitesimally small volumes, yet larger than both the inter-particle separation and the Debye length, ensuring the charge neutrality assumption holds ($\Sigma n_j q_j = 0$). We assume the fluids are non-relativistic and undergo mass and momentum conserving collisional/exchange reactions. Under these assumptions, the continuity equation for species $j$ is given by:
\begin{align}
\label{eq:cont}
    \partial_{t} \rho_{j} + \nabla \cdot (\rho_{j} {\bf u}_{j} ) =0,
\end{align}
where changes in the equilibrium charge are assumed to occur slowly compared to the gyro frequencies, acceleration timescales, and bulk motion of the fluid. Therefore, these effects are neglected here and in the momentum equation.

The momentum equation for species $j$ can be expressed as:
\begin{align}
\label{eqn:mom} & \frac{\partial (\rho_{j} {\bf u}_{j}) }{\partial t}  + \nabla \cdot \left( \rho_{j} \langle {\bf v}_{j}{\bf v}_{j} \rangle \right) =   \rho_{j} {\bf a}_{{\rm ext},\,j}  \\ 
\nonumber &\ \ +  \frac{q_{j} \rho_{j}}{m_{j}}\left[ {\bf E} + \frac{{\bf u}_{j}}{c} \times {\bf B} \right] + \sum_i \rho_j \omega_{ji} ({\bf u}_{i}-{\bf u}_{j}),
\end{align}
where $\bf E$, $\bf B$ are the electric and magnetic fields respectively. The collisional rate $\omega_{ji}$, describing momentum transfer between species $j$ and $i$, with momentum transfer rate coefficient $\langle \sigma v \rangle_{ji}$ is defined as $
\omega_{ji} \equiv \rho_{i} \langle \sigma v \rangle_{ji}/({m_{j} + m_{i}})$.


\subsection{Equations for the bulk fluid}
\label{sec:solution}


Constructing the total momentum equation for the ``gas'' by summing over all non-dust species, we obtain:
\begin{align}
\label{eqn:mass.tot.gaswodust} 
\partial_{t}\rho_g + &\nabla \cdot (\rho_g {\bf U}_g) = 0, \\
\nonumber \frac{\partial (\rho_g {\bf U}_g)}{\partial t} + &\nabla \cdot \left( \rho_g {\bf U}_g {\bf U}_g \right) = - \nabla \cdot \boldsymbol{\Pi}_g \\
\label{eqn:mom.tot.gaswodust}  & + \rho_g \, {\bf a}_{{\rm ext}, g} + \frac{{\bf J} \times {\bf B}}{c} - {\bf F}_{d}, 
\end{align}
where $\rho_g \equiv \sum_{j \neq d} \rho_j$ denotes the gas density. Similarly, the terms ${\bf a}_{{\rm ext},g}$ and ${\bf U}_g$ are defined as mass-weighted averages over all gas species. The gas pressure tensor, \(\boldsymbol{\Pi}_g\), is given by \(\sum_{j \neq d} \rho_{j} \langle {\bf v}_{j} {\bf v}_{j} \rangle - \rho_{j} {\bf U}_g {\bf U}_g\). The current density \({\bf J}\), defined as \(\sum_{j} n_{j} q_{j} \delta {\bf u}_{j,g}\), includes contributions from dust as well, ensuring that it satisfies Ampere's law. ${\bf F}_{d}$ represents the ``back-reaction'' force exerted on the dust by the gas. This arises from Lorentz forces and collisional/drag interactions, giving:
\begin{align}
\nonumber {\bf F}_{d} &\equiv \sum_{i \in d} \frac{q_{i} \rho_{i}}{m_{i}}\left[  {\bf E}^{\prime} + \frac{\delta {\bf u}_{d,g}}{c} \times {\bf B} \right]  \\
\label{eqn:Fdust.definition} & + \sum_{i \in d} \sum_{j \in g} \rho_{i}\omega_{ij} (\delta {\bf u}_{j, g} - \delta {\bf u}_{i,g}).
\end{align}
The sum over all dust species $i$, while ${\bf E}^{\prime} \equiv {\bf E} + ({\bf U}_g/c) \times {\bf B}$, and $\delta{\bf u}_{j,g} \equiv {\bf u}_{j} - {{\bf U}_g}$ denote the electric field and the drift velocity in the co-moving mean gas frame, respectively. 

 \subsection{Equations for the individual species}

Reformulating Eq.~\eqref{eqn:mom} in the co-moving frame of the mean gas velocity, we obtain the following for the drift velocity of each gas component:
\begin{align}
\label{eqn:drift.species} \frac{1}{\rho_{j}}D_{t} &{ (\rho_{j} \delta {\bf u}_{j,g})} = \delta{\bf a}_{j,g} - {\bf G}_{j}  + \frac{q_{j}}{m_{j}} {\bf E}^{\prime} +\frac{\bf{F}_d}{\rho_g}\\
\nonumber  &+ \sum_i  \omega_{ji} (\delta{\bf u}_{i,g}-\delta{\bf u}_{j,g}) + \left( \frac{q_{j}}{m_{j}} \delta {\bf u}_{j,g} - \frac{{\bf J}}{\rho_g} \right) \times \frac{\bf B}{c} ,
 \\ 
\nonumber &\quad \quad{\bf G}_{j} \equiv \frac{1}{\rho_{j}} \nabla \cdot \boldsymbol{\Pi}^{\prime}_{j} - \frac{1}{\rho} \nabla \cdot \boldsymbol{\Pi}_g , 
\end{align}
where $\boldsymbol{\Pi}_{j}^{\prime} \equiv  \rho_{j} \langle \delta{\bf v}_{j} \delta{\bf v}_{j}\rangle + \rho_{j} \delta{\bf u}_{j,g} \delta {\bf u}_{j,g}$, Here, $\delta {\bf{a}}_{j,g} \equiv {\bf a}_{{\rm ext},j} - {\bf a}_{{\rm ext}, g}$ denotes the deviation of the external acceleration on species $j$ from the gas-mass-weighted mean external acceleration on the gas, and $\delta {\bf v}_{j} \equiv  {\bf v}_{j} - {\bf u}_{j}$ is the velocity dispersion. We define a useful total derivative $D_{t} {\bf X} \equiv \partial_{t} {\bf X} +  {\bf U} \cdot \nabla {\bf X}$.

Note that the ${\bf G}_{j}$ term generalizes ``Biermann-like'' battery terms,  while the $\delta{\bf a}_{j, g}$ term extends to batteries driven by external acceleration often through radiation pressure effects. We are interested in cases where radiative battery effects are much stronger than the Biermann battery, i.e. when $|\nabla \cdot \rho_j \delta {\bf{v}_j} \delta {\bf{v}_j}| \sim |\nabla \cdot \rho_j \delta {{\bf{u}}_{j,g}}\delta {{\bf{u}}_{j,g}}| \sim |{\bf G}_{j}| \ll |\delta{\bf a}_{j,g}|$, so will make this approximation below and justify it subsequently \citep[see][for additional conditions where the pressure tensor terms are typically regarded as negligible]{wardle.ng:1999.nonideal.coefficients, tassis.mouschovias:2005.nonideal.formulation.review.accretion, tassis.mouschovias:2007.nonideal.mhd.clouds.formulations.analytic.models}. Considering radiation pressure as the principal driver of external acceleration forces already suggests that we should prioritize the contributions from species with high cross-sections for radiation, such as dust, which we will demonstrate below.

\begin{figure*}
    \centering
    \includegraphics[width=1\linewidth]{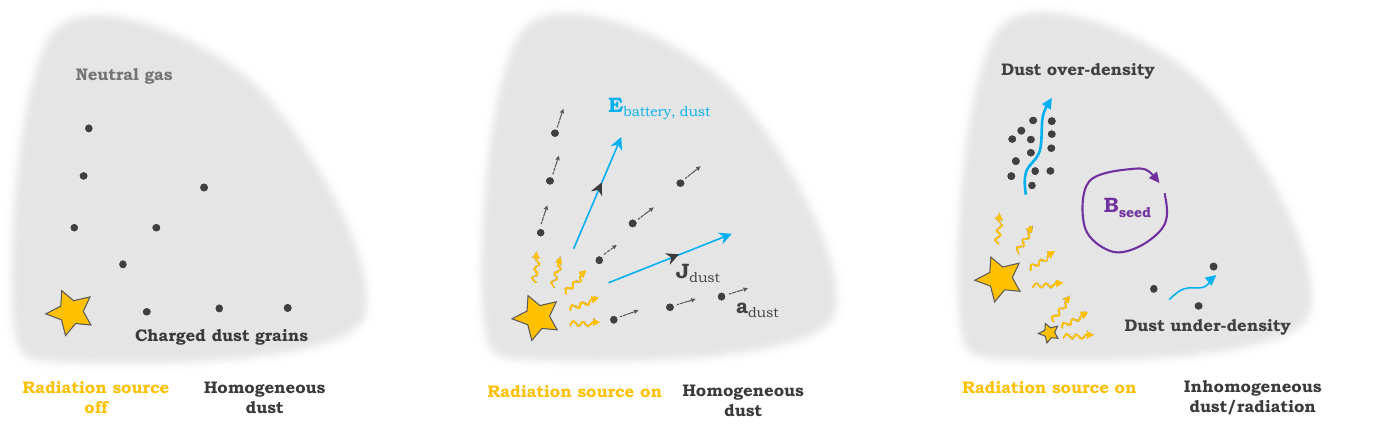}
    \caption{Cartoon representation of the dust radiative battery mechanism for magnetic field generation. In the leftmost panel, with no external radiation, the system consists of neutral gas, electrons, ions, and charged dust grains in a homogenous distribution, resulting in no induced electric field. Upon introducing a radiation source, the charged dust grains experience acceleration, generating a dust current and producing an electric field due to charge separation. In regions with spatial fluctuations in the dust distribution, a non-zero curl of the electric field develops, leading to the generation of a seed magnetic field through the dust battery mechanism.}
    \label{fig:cartoon}
\end{figure*}

\subsection{Solution for arbitrary dust current}
\label{sec:solution.arbitrary.dust.vel}

Even with Amperes law to specify ${\bf J} = (c/4\pi) \nabla \times {\bf B}$, local charge neutrality, the definitions of ${\bf U}_{g}$ above, and known values of $\rho_{j}$, $\omega_{ji}$, etc., Eq.~\eqref{eqn:mom.tot.gaswodust} and the other expressions above do not close, requiring some closure for the microscopic pressure tensors plus expressions for $\delta {\bf u}_{j,\,g}$ and ${\bf E}^{\prime}$. The usual magnetohydrodynamic (MHD) approximation resolves this by assuming a scale hierarchy, where the electromotive and gyro frequencies (terms proportional to the charge-to-mass ratio $q_{j}/m_{j}$) in Eq.~\eqref{eqn:drift.species} are much faster than other terms, and so the charged-species drifts come into equilibrium much faster than the timescales for evolution of ${\bf U}_{g}$ on macroscopically large scales (e.g.\ the macroscopic gradient length scales on large ISM scales; \citealt{nakano:1986.nonideal.mhd.formulation.review, wardle.ng:1999.nonideal.coefficients, tassis.mouschovias:2005.nonideal.formulation.review.accretion,tassis.mouschovias:2007.nonideal.mhd.clouds.formulations.analytic.models}).

Dust grains, having orders-of-magnitude lower charge-to-mass ratio compared to ions and electrons, have much slower gyro frequencies and would reach equilibrium on much longer time (and spatial) scales, potentially extending to macroscopic scales (depending on the grain properties). So first, we can calculate ${\bf E}^{\prime}$ by making the standard MHD assumption for the free electrons and ions (the electrons come into equilibrium first, having the largest charge-to-mass ratio), but still allowing for an arbitrary dust distribution function. The solution for ${\bf{E}}^\prime$ can then be combined with the induction equation, $\partial_{t} {\bf B} = -c \nabla \times {\bf E}^{\prime}$, to compute the evolution of the magnetic field. Specifically, the non-zero curl of the electric field, as derived from the system dynamics, serves as the source term for magnetic field generation. 


Completely general expressions here are again not particularly instructive or helpful, but we show below that the limit where we expect the dust battery to be important is when the gas is mostly neutral. So it is helpful to take the limit of large neutral density, specifically assuming that $\rho_{g}\approx \rho_{n} \gg \rho_{j}$ for all charged $j$, and that the collision rates between electrons/ions and neutrals dominate over electron-electron or ion-ion collisions or charge exchange reactions\footnote{By definition of ${\bf U}_{g}$, $\delta {\bf u}_{n,\,g} = -\rho_{n}^{-1} \sum_{j \in g} \rho_{j} \delta {\bf u}_{j,g} \approx \boldsymbol{0}$ in this limit.}. This is well-justified for the regimes of interest \citep{nakano:1986.nonideal.mhd.formulation.review,wardle.ng:1999.nonideal.coefficients}. As noted above we are also interested by definition in the cases where the Biermann-like battery term ${\bf G}$ is small, so will neglect it as well. Then Eq.~\eqref{eqn:drift.species} for the charged gas species (electrons, ions) becomes: 
\begin{align}
\label{eq:eom}
    -\delta{\bf{a}}_{j,g} \approx \frac{q_{j}}{m_{j}} {\bf E}^{\prime} + \Omega_{j} \delta {\bf u}_{j,g} \times \bhat - \omega_{jn} (\delta {\bf u}_{j,g}-\delta {\bf u}_{n,g}),
\end{align}
with $\Omega_{j} \equiv q_{j} B/m_{j} c$ being the signed cyclotron frequency. We will assume that the radiative acceleration term, $\delta{\bf{a}}_j$, is small compared to the terms on the right-hand side for ions and electrons—a simplification that will be justified later when deriving the electric field induced by the battery effect.

The total current ${\bf J}$ can be expressed as the sum of the currents from individual species:
\begin{align}
    {\bf J} - {\bf J}_{d} = \sum_{j \in g} n_{j} q_{j} \delta {\bf u}_{j,\,g} = n_{+} q_{+} \delta {\bf u}_{+,g} + n_{e} q_{e} \delta {\bf u}_{e,g},
\end{align}
where the latter expression takes our four-species limit, and ${\bf J}_{d} \equiv \sum_{i \in d} n_{i} q_{i} \delta {\bf u}_{i}$ is the ``dust current.'' 

In this limit, the resulting \({\bf E}^{\prime}\) is identical to the usual non-ideal MHD solution \citep{cowling:1976.mhd.book,ichimaru:1978.nonideal.mhd.deriv.and.assumptions,nakano:1986.nonideal.mhd.formulation.review}, but with (1) the modified ``effective current'' \({\bf J} - {\bf J}_{d}\) and (2) with the modified ``effective charge balance'' $\sum_{j \in g} n_{j} q_{j} = -\sum_{i \in d} n_{i} q_{i}$ or \(n_{+} q_{+} + n_{e} q_{e} = -n_{d} q_{d}\). If the collision coefficients \(\omega_{ij}\) are not strong functions of the drift speeds themselves, the equations above form a linear system, and \({\bf E}^{\prime} \rightarrow {\bf E}^{\prime}_{J} + {\bf E}^{\prime}_{{\rm bat},\,d}\) can be decomposed into the usual non-ideal (Ohmic, Hall, ambipolar) terms in \({\bf E}^{\prime}_{J}\) (computed with the given species abundances $n_{j}$ by setting ${\bf J}_{d}\rightarrow \boldsymbol{0}$) and a dust battery term \({\bf E}^{\prime}_{{\rm bat},\,d}\) computed by taking \({\bf J} \rightarrow \boldsymbol{0}\). Finally, it is convenient to express \({\bf E}^{\prime}_{{\rm bat}, d}\) using the basis vectors \(\{{\bf J}_{d}, {\bf J}_{d} \times \hat{\bf{B}}, {\bf J}_{d} \times \hat{\bf{B}} \times \hat{\bf{B}}\}\), with coefficients analogous to the Ohmic, Hall, and ambipolar terms, \(\alpha_{O}\), \(\alpha_{H}\), and \(\alpha_{A}\), as
\begin{align}
\label{eqn:Ebat.general}
{\bf E}^{\prime}_{{\rm bat}, d} &= -\alpha_{O} {\bf J}_{d} - \alpha_{H} {\bf J}_{d} \times \bhat - \alpha_{A} {\bf J}_{d} \times \bhat \times \bhat\ .
\end{align}

We obtain the closed-form analytic expressions: 
\begin{align}
\label{eqn:alpha.o} \alpha_{O} &\equiv \frac{\omega_{en} \omega_{+n}}{\mu_{+} \omega_{en} + \mu_{e} \omega_{+n}},\\
\nonumber \mu_{j} &\equiv \frac{n_{j} q_{j}^{2}}{m_{j}} = \frac{\omega_{{\rm plasma},\,j}^{2}}{4\pi}, \\
\label{eqn:alpha.h} \alpha_{H} &= \frac{[\mu_{+} \Omega_{+} (\omega_{en}^{2} + \Omega_{e}^{2}) + \mu_{e} \Omega_{e} (\omega_{+n}^{2} + \Omega_{+}^{2})]}{\psi_{H}}, \\
\nonumber \psi_{H} &= [\mu_{+} (\omega_{en} - \Omega_{e}) + \mu_{e} (\omega_{+n} - \Omega_{+})] \\
\nonumber & \quad \quad \cdot[\mu_{+} (\omega_{en} + \Omega_{e}) + \mu_{e} (\omega_{+n} + \Omega_{+})], \\
\nonumber \alpha_{A} &= \frac{\mu_{e} \mu_{+} (\Omega_{e} \omega_{+n} + \Omega_{+} \omega_{en})^{2}}{\psi_H \left( \mu_{+} \omega_{en} +\mu_{e} \omega_{+n} \right)} \\
\label{eqn:alpha.a}  & \quad \quad+\frac{2 \Omega_{e} \Omega_{+} (\mu_{+}^{2} \omega_{en}^{2} + \mu_{e}^{2} \omega_{+n}^{2})}{\psi_H \left( \mu_{+} \omega_{en} +\mu_{e} \omega_{+n} \right)},  
 \end{align}
where $\omega_{{\rm plasma}, j}$ is the plasma frequency for species $j$.

In the weak magnetic field limit, $\alpha_{O}$ remains unmodified, while \(\alpha_{H}\) and \(\alpha_{A}\) scale as \(\propto B\) and \(\propto B^{2}\), respectively, similar to their Hall and ambipolar diffusion counterparts, rendering them less significant.  Consequently, the $-\alpha_{O} {\bf J}_{d}$ term constitutes the true ``battery'' term acting, unlike its Ohmic analog, to generate non-zero ${\bf B}$ where there is none, and the generated electric field is, to first order, proportional to the dust current.

In the regime where electrons are highly depleted onto dust grains (\(\mu_e \rightarrow 0\)) -- a condition prevalent in mostly neutral gas (see \citealt{umebayashi1980recombination, umebayashi1983densities, umebayashi1990magnetic, nishi1991magnetic, nakano2002mechanism} and Section \ref{sec:neutrality} for constraints on ionization fraction, density, and temperature) -- we can simplify even further to obtain: 
\begin{align}
\label{eqn:alpha.limits} \alpha_{O} & \rightarrow \omega_{+n}/\mu_{+}\ , \\ 
\nonumber \alpha_{H} & \rightarrow (\Omega_{+}/\mu_{+})\,[(\omega_{en}^{2} + \Omega_{e}^{2})/(\omega_{en}^{2} - \Omega_{e}^{2})]\ , \\ 
\nonumber \alpha_{A} & \rightarrow 2\,(\Omega_{+}/\mu_{+})\,[(\omega_{en} \Omega_{e})/(\omega_{en}^{2} - \Omega_{e}^{2})]\ . 
\end{align}

Physically, while it is obvious that the dust battery ${\bf E}^{\prime}_{{\rm bat},\,d}$ should scale with the dust current ${\bf J}_{d}$, the scaling of the coefficient $\alpha_{O}$ is less intuitive. The key is that a battery effect relies on generating charge {\em separation}, which is captured by $\alpha_{O} \sim \omega_{jn}/\mu_{j} \propto (m_{j} \omega_{jn}) / n_{j} q_{j}^{2}$. This is just the inverse of the usual ``mobility'' parameter in mostly-neutral systems \citep{ichimaru:1978.nonideal.mhd.deriv.and.assumptions, nakano:1986.nonideal.mhd.formulation.review}: if the charged gas species are infinitely mobile (e.g.\ massless and collisionless) they would be dragged perfectly with the dust ($\alpha_{O}\rightarrow0$), preventing any battery. Collisions and finite mass reduce the dust mobility, enabling greater charge separation and hence a stronger battery.

With these expressions, we can validate that our earlier assumption that the acceleration terms for ions and electrons are negligible—holds when \(|\delta{\bf{a}}_{i,g}| \ll |\delta{\bf{a}}_{e,g}| \ll |n_d Z_d \langle \sigma v\rangle_{en} \delta{\bf{u}}_{d,g}|\, n_n/n_e\). This condition is easily met in high neutral density environments and when electrons are mostly depleted onto dust grains \citep{umebayashi1980recombination, umebayashi1983densities, umebayashi1990magnetic, nishi1991magnetic, nakano2002mechanism}.

Additionally, we find that the dust battery term will be non-negligible compared to standard Ohmic resistivity if \(|{\bf J}_{d}| \gtrsim m_{e}/m_{+} |{\bf J}|\), a condition easily satisfied for the weak magnetic-fields regimes of interest. Estimates for the magnetic field strength and scales at which Ohmic resistivity becomes comparable to the battery are discussed in \S \ref{sec:saturation}.

Another key limiting case is the fully ionized regime. Solving Eq.~\eqref{eqn:drift.species} in the limit $n_n \rightarrow 0$ and considering collisions exclusively between charged species, the Ohmic, Hall, and ambipolar diffusion terms simplify to the following forms:
\begin{align}
\label{eqn:alpha.limits1} \alpha_{O} &\rightarrow \frac{m_{e}(m_{+} n_{+}  + m_{e}n_{e})(n_{d}q_{d}\omega_{e+}-n_{+}q_{+}\omega_{ed})}{n_{+} n_{d} q_{d} n_{e} q_{e} (m_{+} q_{e} - m_{e} q_{+})} , \\
\nonumber \alpha_{H} &\rightarrow \frac{\Omega_{-}}{\mu_{-}} , \\
\nonumber \alpha_{A} &\rightarrow 0 .
\end{align}

Using the definition of $\omega_{ij}$ and considering the limits $m_e \ll m_+$, $n_d \ll n_e \sim Z_+ n_+$, $q_d \sim \pm Z_d q_e $, and $q_+ \sim - Z_+ q_e$, we can simplify the Ohmic term as follows: $\alpha_{O} \sim \frac{m_e}{q_e^2} \left(\frac{\langle \sigma v \rangle_{+e}}{Z_+} \pm \frac{\langle \sigma v \rangle _{ed}}{Z_d}\right)$. This matches the intuition from the mostly neutral case. In the fully ionized regime, ions dominate the fluid inertia due to their mass hierarchy, similar to how neutrals dominated in earlier limits. The much shorter response time of electrons compared to protons makes it more efficient to balance the dust current with an electron counter-current rather than a proton one. Consequently, charge separation is controlled by electron-ion and electron-dust collisional rates, or equivalently, by electron mobility. As we will later demonstrate in \S \ref{sec:solutions.ionized}, using assuming reasonable parameters, this mechanism can generate an electric field in fully ionized environments comparable in magnitude to that found in the mostly neutral limit.

Eq. ~\eqref{eqn:Ebat.general} plus the induction equation $\partial_{t} {\bf B} = -c \nabla \times {\bf E}$ allows one to compute the dust battery given knowledge of an arbitrary ${\bf J}_{d}$. As such, it can be directly implemented into any particle-in-cell (PIC) or MHD-PIC code which already follows the dust grains (computing ${\bf J}_{d}$ as specified above locally as a sum over dust particles/species). Examples include \citet{seligman:2018.mhd.rdi.sims, hopkins:2019.mhd.rdi.periodic.box.sims}.


A cartoon illustrating this mechanism and the generation of the magnetic field from a non-zero \({\partial{\bf B}}/{\partial t} = - c \, (\nabla \times {\bf E}_{{\rm bat}, d}^\prime)\) is presented in Figure \ref{fig:cartoon}.

\subsection{Derivation of the dust current in the terminal velocity limit}
\label{sec:tva}

To make further analytic progress toward estimating the battery strength, we require an analytic expression for ${\bf J}_{d}$. First consider a simple heuristic scaling for intuition using the usual  approximation for aerodynamic grains in a mostly-neutral gas with just Epstein drag \citep{nakagawa1986settling}. Grains experiencing some acceleration ${\bf a}_{d, g}$ from e.g.\ radiation are slowed by collisions/drag with neutrals ${\bf a}_{\rm drag} \sim -\omega_{dn} \delta {\bf u}_{d, g}$. This gives a ``terminal velocity'' $\delta {\bf u}_{d, g} \sim {\bf a}_{d, g}/\omega_{dn}$ hence the battery ${\bf E}^{\prime}_{{\rm bat},\,d} = -\alpha_{O} {\bf J}_{d} \sim -\alpha_{O} n_{d} q_{d} {\bf a}_{d, g} / \omega_{dn}$}. We will see this simple expression validated below via a more complete calculation.


More rigorously, we revisit Eqs.~\eqref{eq:cont}-\eqref{eqn:drift.species} and apply the same assumptions to derive Eq.~\eqref{eq:eom}, while retaining all “cross-collision” terms for completeness. Following our approach in \S~\ref{sec:solution.arbitrary.dust.vel}, we now specify to our four-component system, but instead of just solving Eq.\eqref{eqn:eom.all} for electrons and ions in terms of an arbitrary dust current $\delta {\bf u}_{d}$, we assume that the dust species also reach their local terminal velocities. We obtain that the gas and dust components obey:
\begin{align}
\label{eqn:eom.all}
      - \delta {\bf{a}}_{j,g}  =  \frac{q_{j}}{m_{j}} {\bf E}^{\prime} + \Omega_{j} \delta {\bf u}_{j,g} \times \bhat  - \sum_{i} \omega_{ji} ( \delta {\bf u}_{j,g} - \delta {\bf u}_{i,g}  ). 
\end{align}
Once again this is a linear problem, so long as the collision rates $\omega_{ji}$ are weakly dependent on the $\delta {\bf u}_{j}$, and we can decompose ${\bf E}^{\prime}$ into the standard non-ideal plus dust current terms with ${\bf E}^{\prime}_{{\rm bat},\,d}$ written as Eq.~\eqref{eqn:Ebat.general}.\footnote{Note that we can re-derive everything defining the relative velocities \(\delta \mathbf{u}_j\) and accelerations \(\delta \mathbf{a}_j\) with respect to the total fluid, where \(\delta \mathbf{a}_j\) represents the deviation from the total external acceleration \(\mathbf{a}_{\text{ext}} = \sum_{j} \mathbf{a}_{\text{ext},j}\), and \(\delta \mathbf{u}_j\) represents the velocity relative to the system’s bulk velocity \(\mathbf{U} = \sum_{j} \mathbf{u}_j\), inclusive of all species (i.e.\ a single fluid including dust). But this gives identical expressions in what follows, since we are always interested in the limit where the dust mass is a small fraction of the total fluid system mass.}

A complete matrix formulation including perpendicular components is provided in Appendix~\eqref{sec:general.solution.appendix}, but since we are interested in the ``true battery'' terms here we can focus on the parallel component (equivalently, the limit $|{\bf B}|\rightarrow 0$). This gives
\begin{align}
\nonumber {\bf E}^{\prime}_{{\rm bat},d} &\approx -\alpha_{O}  {\bf J}_{d} \approx -m_{d} \delta {\bf a}_{d,g} \left[\frac{m_{e} m_{+} \,\Psi_{1}}{q_{d} m_{e} m_{+} \, \Psi_{1} + \Psi_{2}} \right], \\ 
\nonumber \Psi_{1} &\equiv  - n_{e} q_{e} \omega_{en} \omega_{+d} - n_{+} q_{+} \omega_{ed} \omega_{+n} \\
\nonumber & + n_{d} q_{d} \left( \omega_{e+} \omega_{+n} + \omega_{en} \omega_{+e} + \omega_{en} \omega_{+n} \right), \\ 
\nonumber \Psi_{2} &\equiv   n_{+} q_{+} \left( q_{+} \hat{\psi}_{ed} - q_{e} \hat{\psi}_{+d}  \right) \\
\nonumber & - n_{d} q_{d} \left[ q_{e} \hat{\psi}_{d+} + q_{+}  m_{e} m_{d} \left(\omega_{e+} \omega_{dn} - \omega_{d+} \omega_{en} \right)  \right], \\ 
\label{eq:ebatd} \hat{\psi}_{ij} &\equiv m_{i} m_{j} \left[\omega_{ij} \omega_{jn} + \omega_{in} \sum_{k \ne j} \omega_{jk}  \right].  
\end{align}
This is still somewhat opaque, but in the mostly-neutral limit--specifically, when electron-neutral collisions become more frequent than electron-ion collisions--simplifies greatly to:
\begin{align}
\label{eqn:ebat.neutral}{\bf E}^{\prime}_{{\rm bat},d} &\approx - \frac{m_{d} \delta {\bf a}_{d,g}}{q_{d}} \frac{\mu_{d} \omega_{en} \omega_{+n} }{\mu_{d} \omega_{en} \omega_{+n} + \mu_{e} \omega_{+n} \omega_{dn} + \mu_{+} \omega_{en} \omega_{dn}} 
\end{align}
Taking the same limits as our heuristic derivation above and Eq.~\eqref{eqn:alpha.limits}, the right-hand side of Eq.~\eqref{eq:ebatd} simplifies to $\sim -m_{d} {\bf a}_{d, g} \mu_{d} \omega_{en} \omega_{+n}/q_{d} \mu_{+}\omega_{en} \omega_{dn} \sim -(\omega_{+n}/\mu_{+})\,n_{d} q_{d} ({\bf a}_{d, g}/\omega_{dn}) = -\alpha_{O} {\bf J}_{d}$, corresponding to our simple heuristic expectation.

Another limit of interest is of a fully-ionized system, for which Eq.~\eqref{eq:ebatd} becomes:\footnote{One can re-derive Eq.~\eqref{eqn:ebat.ion} from scratch, or take limits of Eq.~\eqref{eq:ebatd}, but the latter must be done with care to avoid spurious divergences (e.g.\ adding the neutral inertia to ions via $\omega_{+n} \rightarrow \infty$ while simultaneously taking $n_{n} \rightarrow 0$).}
\begin{align}
\label{eqn:ebat.ion}{\bf E}^{\prime}_{{\rm bat},d}\approx \frac{\rho_{e} (n_{+} q_{+} \omega_{ed} - n_{d} q_{d} \omega_{e+}) \,\delta{\bf a}_{d,g}}{\rho_{+} \mu_{+} \omega_{de} + \rho_{e} (\mu_{e} \omega_{d+} + \mu_{d} \omega_{e+})} .
\end{align}

The symmetry of the problem means that we can immediately obtain the radiation-battery effect due to electrons or ions from Eq.~\eqref{eq:ebatd} by exchange of indices $e \leftrightarrow d$ or $+ \leftrightarrow d$. If we further simplify to a fully-ionized, dust-free fluid with infinitely mobile electrons ($n_{n} \rightarrow 0$, $n_{d} \rightarrow 0$, $m_{e}/m_{+}\rightarrow 0$), we obtain the simple electron battery expressions in e.g. \citet{harrison1973magnetic,gopal2005generation,matarrese2005large, ando2010generation}.




\subsubsection{Radiation-driven dust acceleration}
\label{sec:rad.battery.neutral}

Per \S~\eqref{sec:tva}, Eq.~\eqref{eq:ebatd} simplifies to Eq.~\eqref{eqn:ebat.neutral} in the mostly neutral limit. Let us further consider an external acceleration driven by radiation pressure,\footnote{The acceleration $\delta {\bf a}_{d,g}$ does not have to come from radiation, necessarily, and other mechanisms to introduce an effective $\delta {\bf a}_{d,g}$ (e.g.\ drift of dust in a hydrostatically pressure-supported gas system) can be important in some regimes (reviewed in \citet{hopkins2018resonant}). But the cases of greatest interest, where $\delta {\bf a}_{d,g}$ is large, often owe to radiation driving, and it is convenient to compare to the more well-studied radiation-electron battery.} an extremely common astrophysical situation \citep{elitzur2001dusty, krumholz2013numerical, zhang2017radiation, steinwandel2022optical,hopkins2022dust, soliman2023dust}. In this case,  $\delta {\bf a}_{d,g} \approx \sigma_{d,r} {\bf F}_{\rm rad}/m_{d} c$ in terms of the incident radiation flux ${\bf F}_{\rm rad}$ and effective cross-section for radiation absorption plus scattering. \footnote{Since we are interested in general scalings, we ignore the subtleties of e.g.\ non-ionizing absorption versus scattering and anisotropic scattering.}  Recall that $\delta{\bf a}_{j,g}$ is the acceleration of species $j$ relative to the mean acceleration acting directly on the gas species, so generally $\delta {\bf a}_{d,g} \approx {\bf a}_{d,g}$, but $\delta {\bf a}_{e,g}$ will be smaller in magnitude than ${\bf a}_{e,g}$. With this, plus the definitions $\langle \sigma v \rangle_{ij} = \sigma_{ij} v^{\rm eff}_{ij}$ where $v^{\rm eff}_{ij}$ can be approximated for our simple purposes by the thermal velocity of the lighter species, we can compare the characteristic strengths of the dust and electron radiation batteries: 
\begin{align}
\label{eq:rationeutral}
\frac{|{\bf E}^{\prime}_{{\rm bat}, d}|}{|{\bf E}^{\prime}_{{\rm bat}, e}|} &\rightarrow \frac{|n_d q_d \omega_{en} \delta {\bf a}_{d,g} |}{ |n_e q_e \omega_{dn} \delta {\bf a}_{e,g} |} \\\
\nonumber&\approx \left|\frac{\sigma_{d,r} {\bf F}_{\rm rad}}{m_{d} c }\right|\left|\frac{m_{e} c }{\sigma_{e,r} {\bf F}_{\rm rad}} \right|
\left[ \frac{ m_{d} \langle \sigma v\rangle_{en} n_{d} |q_{d}| }{ m_{n} \langle \sigma v\rangle_{dn} n_{e} |q_{e}|  } \right]  \\ 
\nonumber  &\sim \frac{\sigma_{en}}{\sigma_{T}} \sqrt{\frac{m_{e}}{m_{n}}} \frac{n_{d}}{n_{e}} \frac{|q_{d}|}{|q_{e}|} \sim 10^{8}\,\frac{n_{d} |q_{d}|}{n_{e}|q_{e}|} 
\end{align}
In the latter we have used $\sigma_{r,d} \sim \sigma_{T}$ (Thompson), $\sigma_{dn} \sim \sigma_{d,r} \sim \sigma_{d}$ (of order the geometric cross-section), and $\sigma_{en} \sim \times 10^{-15} \rm cm^2$ \citep{spitzer1953transport, pinto.galli:2008.momentum.transfer.coefficients.for.weakly.ionized.systems}. 

Note that this ratio depends solely on the charge density ratio between dust grains and free electrons, and is not explicitly dependent on specific grain properties such as size or composition. However, the charging of the grains and their number density may still be influenced by their size. In predominantly neutral environments, where $n_d |q_d| \gg n_e |q_e|$  due to the depletion of electrons onto the grains (see \S~\eqref{sec:neutrality}) \citep{umebayashi1980recombination,umebayashi1983densities, umebayashi1990magnetic, nishi1991magnetic}, the fields generated by the dust are many orders of magnitude greater than those from electrons.

Returning to the battery strength ${\bf E}^{\prime}_{{\rm bat},d}$ in Eq.~\eqref{eqn:ebat.neutral}, the three terms in the denominator dominate in three different limits. If free electrons are not strongly depleted onto dust (the lowest densities/higher ionization fractions, though truly well-ionized cases are discussed in \S~\eqref{sec:solutions.ionized}), the $\mu_{e}$ term dominates.\footnote{Caution is needed in this lower-density, higher ionization limit, however, since at relatively low temperatures $T\ll 10^{5}$\,K, the Spitzer collision rate $\omega_{e+}$ can then be larger than the neutral collision rates, and the full Eq.~\eqref{eq:ebatd} is needed. But in the limit where $\omega_{e+}$ becomes the largest frequency while neutrals still dominate the total density, ${\bf E}^{\prime}_{{\rm bat},d} \rightarrow -m_{d} \delta {\bf a}_{d,g}/q_{d}$, identical to the case at the lowest ionization fractions when both electrons and ions are depleted onto dust. In the fully ionized case, this solution remains valid as long as $\omega_{e+}$ continues to dominate the collision frequencies.} If free electrons are depleted onto dust but the positive charge density remains primarily in ions/gas, the $\mu_{+}$ term dominates (the intermediate regime). If the ions are also depleted onto dust (the highest densities/lowest ionization fractions; see \citealt{umebayashi1980recombination, desch2001magnetic, nakano2002mechanism}), the $\mu_{d}$ term dominates. In the latter limit, we obtain the extremely simple expression $|{\bf E}^{\prime}_{{\rm bat},d}| \rightarrow m_{d} \delta {\bf a}_{d,g} / q_{d} \sim {\bf F}_{\rm rad} \sigma_{d,r}/ q_{d} c \sim 2 \times 10^{-15} \,{\rm statV\,cm^{-1}} (R_{\rm grain}/{\rm n m})^{2}\,({\bf F}_{\rm rad} / {\rm erg \, cm^{-2} \, s^{-1}})$, where we assume $|q_{d}| \sim |q_e|$ which holds for typical dust grain sizes under conditions of low temperature and/or low ionization fractions.

More generally we can estimate
\begin{align}
{{\bf E}^{\prime}_{{\rm bat}, d}}& \sim \frac{{\bf F}_{\rm rad} \sigma_{+n}}{q_{+} c} \frac{n_{d} q_{d} }{n_{+} q_{+} (1 + \Psi_{3} )}\\
\nonumber  &\sim \frac{{\bf F}_{\rm rad}\sigma_{+n}}{q_e c} \sim 2 \times 10^{-15}\, {\rm statV\,cm^{-1}}\, \frac{{\bf F}_{\rm rad}}{\rm erg\,s^{-1}\,cm^{-2}}, \\ 
\nonumber 
\Psi_{3} &\approx \frac{n_{e}}{n_{+}} \frac{\sigma_{+n}}{\sigma_{en}} \sqrt{\frac{m_{n}}{m_{e}}} + \frac{n_{d}}{n_{+}}\frac{\sigma_{+n}}{\sigma_{dn}} \ .
\end{align}
In the numerical expression we assume $\Psi_{3}$ is small, corresponding to electrons being depleted onto grains (implying $|n_{d} q_{d}| \approx |n_{+} q_{+}|$), though this term captures the corrections for the different regimes above. 

As one would expect, ${{\bf E}^{\prime}_{{\rm bat}, d}} \propto {\bf F}_{\rm rad}$. However, in this regime, as previously discussed, the electric field induced by the dust battery is independent of the specific characteristics or quantity of the dust, owing to the condition \(n_e \ll n_d Z_d\). Instead, the electric field is primarily influenced by the ion-neutral collision cross-section. This occurs because a larger collision cross-section implies stronger coupling between ions and neutrals, leading to a greater lag of ions relative to the dust within the acceleration field.

To estimate the magnetic field seeding rate, we apply the induction equation to the radiation field of a source with luminosity \( L \) at a distance \( R \), where the flux is \(|{\bf F}_{\rm rad}| \sim L / 4 \pi R^2 \). Assuming an electric field gradient scale, or equivalently a radiation field gradient of \( \nabla \sim 1 / \ell \), reflecting fluctuations in dust density or the source itself on a scale $\ell$. These fluctuations have been shown to be significant across scales ranging from $\lesssim$\, AU to $\gg {\rm kpc}$ \citep{abergel:2002.size.segregation.effects.seen.in.orion.small.dust.abundances, miville2002isocam,gordon:2003.large.variations.extinction.curves.in.lmc.smc.mw.sightlines,youdin2004particle, yoshimoto2007self, flagey:2009.taurus.large.small.to.large.dust.abundance.variations,paradis2009spatial, pan2011turbulent, boogert2013infrared,  ysard:2005.dust.extinction.curve.variations.diffuse.ism, ohashi2019radial}. Consequently, we obtain:
\begin{align}
\label{eq:seeding}
\frac{\partial {\bf B}}{\partial t} &= -c\nabla \times { \bf E'_{\text{bat},d}} \\
&\nonumber \sim \frac{0.2\, {\rm m G}}{\rm yr} \left(\frac{L}{L_{\odot}}\right) \left(\frac{\rm AU}{R} \right)^{3} \left(\frac{R}{\ell}\right) \\ 
\nonumber &\sim \frac{4\, {\rm mG}}{\rm yr} \left(\frac{L}{10^{12} L_{\odot}}\right) \left(\frac{\rm pc}{R}\right)^{2} \left(\frac{\rm AU}{\ell} \right)
\end{align}
corresponding to the seeding rates predicted around a sun-like star or a bright AGN in the second and third lines, respectively. These are enormous rates, and there is nothing in principle that restricts this mechanism to plasma ``micro-scales''; it can operate efficiently even for astrophysically large $\ell$, although seeding rates will be correspondingly lower due to the large $\ell$. Again, any other source of differential dust-gas forces $\delta {\bf a}_{d,g}$ of similar magnitude will act in the same manner (e.g.\ drift/settling in disks or dusty atmospheres). This implies generation of non-linearly interesting magnetic fields on coherence scales $\ell$ well within the inertial range of ISM turbulence or other classical dynamo effects (\S~\ref{sec:intro}), on timescales short compared to the local dynamical time.

\subsubsection{Potential Role of Instabilities}
\label{sec:instabilities}

It is worth noting that even if one could contrive a homogeneous, spherically-symmetric medium and radiation source pushing on dust, such that $\nabla \times {\bf E}_{{\rm bat},\,d}^{\prime}$ vanishes, this situation is unstable to the the super-family of resonant drag instabilities (RDIs; \citealt{squire.hopkins:RDI}). These instabilities will amplify fluctuations with a growing curl of ${\bf J}_{d}$ (and therefore ${\bf E}_{{\rm bat},d}^{\prime}$; see e.g. \citealt[][]{moseley2019non}). They manifest across all wavelength/scales $\ell$ larger than the ion gyro radii up to galactic scales or larger\citep{hopkins:2018.mhd.rdi} with growth rates that depend only weakly on the dust mass \citep{squire.hopkins:RDI}. These instabilities arise in both dust consisting of a single grain size and in a spectrum of grain sizes \citep{squire2022acoustic}. Thus, they persumably always generate non-zero battery terms on scales comparable to the inertial range of turbulence and other dynamo effects. They may even act as such directly, filling an analogous role to the Weibel instability in weakly-collisional plasmas (see \S~\ref{sec:intro}), though more detailed modeling of this regime would clearly require the sort of MHD-PIC simulations discussed in \S~\ref{sec:general.equations}. 


\subsubsection{Dissipative Effects}
\label{sec:saturation}

Our calculation indicates that the dust battery mechanism could very efficiently generate strong seed magnetic fields around stars and bright sources in the early Universe. However, it is crucial to consider when this field generation mechanism would saturate due to dissipative effects like Ohmic resistivity, especially considering that we have invoked mostly-neutral environments (where resistive effects become important; see \citealt{ichimaru:1978.nonideal.mhd.deriv.and.assumptions, umebayashi1983densities, nakano1984contraction, nakano:1986.nonideal.mhd.formulation.review}). To estimate this, we can compare (in the mostly neutral limit) the growth rate $\sim c \nabla \times {\bf E}_{{\rm bat},d}^{\prime}$ to the Ohmic dissipation rate $\sim c \nabla \times {\bf E}_{O} \sim -\nabla \times [\eta_{O} (\nabla \times {\bf B})]$ with $\eta_{O} \sim c^{2} \omega_{en} /(4\pi \mu_{e})$ \citep{blaes1994local}. Equating the two implies saturation at field strengths 
\begin{align}
    \label{eqn:bsat}|{\bf B}_{\rm sat}| &\gtrsim \frac{4 \pi  \ell q_e |{\bf F}_{\rm rad}| }{c^2\sqrt{m_e k_B T}} \left ( \frac{\sigma_{+n}}{\sigma_{en}}\right ) \left ( \frac{n_e}{n_n}\right ) \\
    \nonumber &\sim 3 \text{mG} \left(\frac{L}{L_{\odot}}\right) \left(\frac{\rm AU}{R} \right) \left(\frac{\ell}{R}\right)  \left ( \frac{n_e/n_n}{10^{-15}}\right ) \left ( \frac{10 \text{K}}{T}\right )^{1/2}
\end{align}
where $\ell$ is the gradient scale-length and we have inserted the same expressions for $F_{\rm rad}$ from above. The momentum transfer cross-sections are given by $\sigma_{+n} \sim 1.17 \times 10^{-9}{\rm cm^{3}s^{-1}} /|{\bf{v}}_p|$, and 
$\sigma_{+n} \sim 1.97 \times 10^{-9} {\rm cm^{3}s^{-1}}/|{\bf{v}}_e|$ \citep{pinto.galli:2008.momentum.transfer.coefficients.for.weakly.ionized.systems}.

\begin{figure}[htbp]
    \centering
\includegraphics[width=\columnwidth]{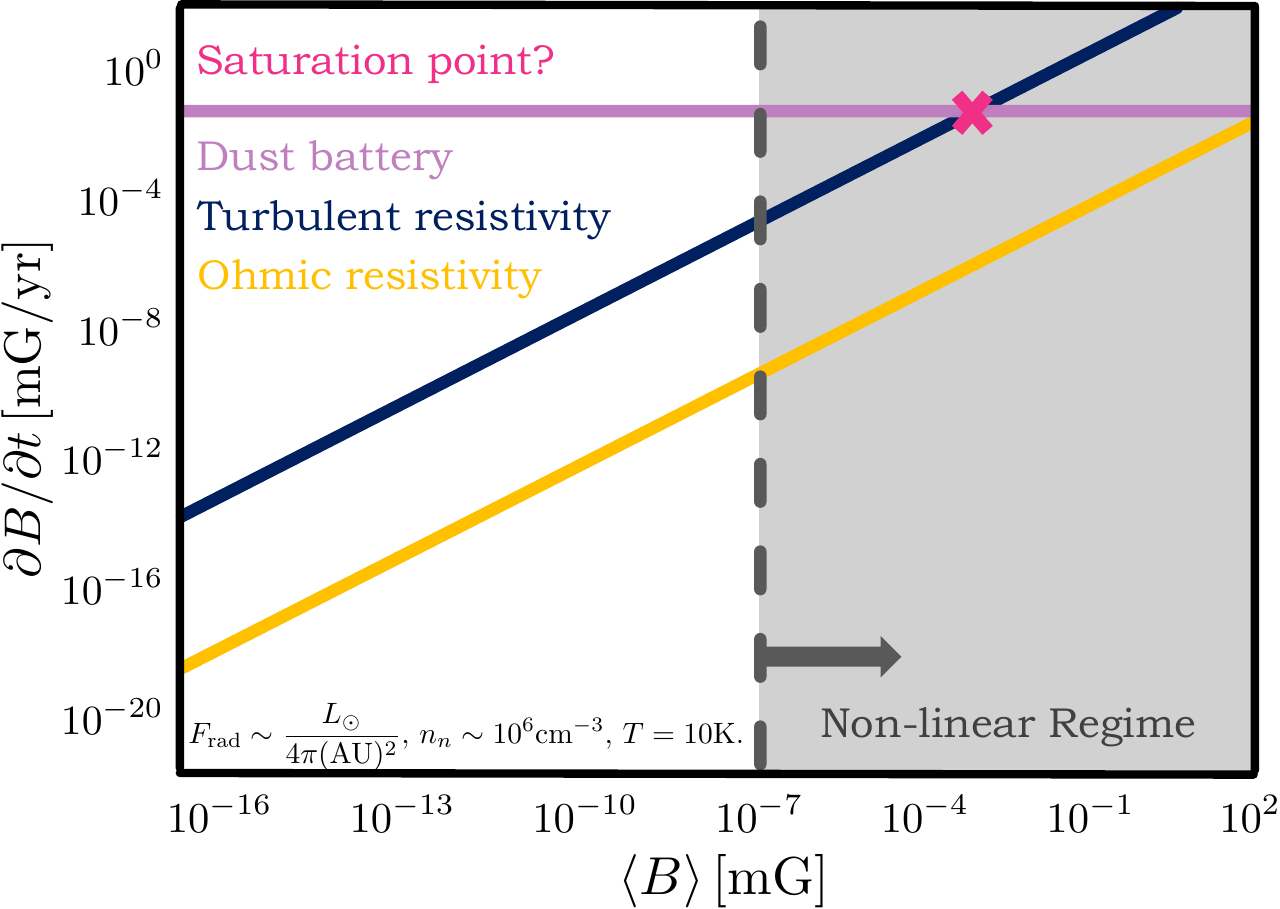}
    \caption{This analysis compares the dust battery seeding rate with both turbulent and Ohmic resistivity rates. The intersection points suggest a saturation field strength of $\langle B \rangle \sim \rm mG$, assuming the seeding rate is counteracted by turbulence or resistive dissipation. These values are derived using the typical parameters outlined in section \ref{sec:saturation}. However, as shown in the figure, the saturation point falls within the non-linear magnetohydrodynamics regime, highlighting the need for simulations to accurately constrain the mean saturation field strength.}
    \label{fig:sat}
\end{figure}

Note that invoking some turbulent resistivity $\partial_{t} {\bf B} \sim - \eta_{\rm turb} {\bf B}/\ell^2$ with $\eta_{\rm turb}(\ell) \sim  \ell v_{\rm turb}(\ell)$ also does not strongly limit $|{\bf B}|$: comparing with Eq.~\eqref{eq:seeding} gives $|{\bf B}_{\rm sat}| \sim 3 \, {\rm \mu G} (L/L_{\odot})\,(R/{\rm AU})^{-2}\,(v_{\rm turb}[\ell]/{300\, \rm km\,s^{-1}})^{-1}$. Here, we assume that the turbulent velocity $v_{\rm turb}$ is comparable to the dust drift velocity in a gas with a number density of $n_n \sim 10^6 \rm cm^{-3}$.  Figure \ref{fig:sat} illustrates these arguments showing the expected saturation magnetic fields set by turbulence and Ohmic resistivity. Note that the saturation amplitude is extrapolated into the nonlinear regime, which we define as the mean magnetic field strength at which the nonlinear terms in Eq.~\eqref{eq:ebatd}—specifically, the Hall-like and ambipolar-like terms, reach 10 \% of the Ohmic-like term.

Therefore, we find that even in environments with very low electron fractions, the dust battery mechanism saturates at fairly high magnetic field strengths on all scales larger than the smallest micro-scales. And if anything this only limits the {\em smallest} wavelength modes, which are least interesting from a cosmological point of view. Consequently, the generation of seed fields, which only need to be orders of magnitude smaller than the computed values, can proceed without substantial dissipation from Ohmic resistivity. While ambipolar diffusion and the Hall effect will affect plasma dynamics and interact with the dust battery mechanism, they will only be important at larger field strengths. Consequently, their effects are not considered in this analysis.

The predicted strength of the seed field and its high expected saturation strength raise important questions regarding the role of dynamos in amplifying fields seeded by dust batteries. Traditionally, seed magnetic fields are considered weak, necessitating amplification via magnetohydrodynamic (MHD) dynamos to reach observed field strengths. However, simulations indicate that when the seed field is sufficiently strong to be dynamically relevant, dynamo amplification saturates early \citep{marinacci2015effects}. In such cases, the magnetic field intensity is primarily determined by flux conservation rather than turbulent motions. Moreover, despite variations in the initial seed field strength, the final magnetic field strength in simulated galaxy halos is generally observed to become relatively uniform, as demonstrated by \citet{marinacci2015large, garaldi2021magnetogenesis}.

 \subsubsection{Solution in the fully ionized limit}
\label{sec:solutions.ionized}

Now consider the opposite regime of a fully-ionized plasma. In this case, Eq.~\eqref{eqn:ebat.ion} simplifies to:
\begin{align}
\label{eq:ratiocomp}
\frac{|{\bf E}^{\prime}_{{\rm bat}, d}|}{|{\bf E}^{\prime}_{{\rm bat}, e}|} &= \frac{|\delta {\bf a}_{d,g}|}{|\delta {\bf a}_{e,g}|}  \frac{ |n_{+} q_{+} \omega_{ed} - n_{d} q_{d} \omega_{e+}|}{|n_{+} q_{+} \omega_{d-} - n_{e} q_{e} \omega_{d+}|} \\ 
\nonumber &\sim \frac{m_{e} \sigma_{d,r}}{m_{d} \sigma_{T}}  \frac{ |n_{+} q_{+} \omega_{ed} - n_{d} q_{d} \omega_{e+} |}{ |n_{e} q_{e} \omega_{d+}|} \\ 
\nonumber &\sim \sqrt{\frac{m_{e}}{m_{+}}} \frac{n_{d} \sigma_{d} f_{q}}{n_{e} \sigma_{T}} \left| 1  - \frac{q_{d}}{q_{+}}\frac{\sigma_{e+}}{\sigma_{ed}} \right| \\
\nonumber&\sim 10^{5} \, \frac{|q_{d}|}{|q_{d}^0|} \left ( \frac{ \rm Z}{\rm Z_{\odot}}\right ) \left ( \frac{{1 \rm nm}}{R_{\rm grain}} \right )\,\left( \frac{T}{10^{4}\,{\rm K}} \right)^{-2}.
\end{align}
In the last expression we consider the case for $|q_{d} \sigma_{e+}/q_{+}\sigma_{ed}| \gg 1$ and scale the dust-to-gas ratio with the metallicity $\rho_{d}/\rho \sim 0.01\,Z/Z_{\odot}$ \citep{issa1990dust, james2002scuba, draine2007dust,bendo2010jcmt, magrini2011herschel}. Additionally, we scale $|q_{d}|$ to some ```expected'' charge ${q_{d}^0} \sim -q_e (R_{\rm grain}/\text{nm})^{2}$, since grains in well-ionized environments can be multiply-charged by collisions or photo-electric effects\footnote{Eq.~\eqref{eq:ratiocomp} scales to the field-emission-limited charge, appropriate for these grain sizes and temperatures, $q_{d}/e \sim -(R_{\rm grain}/\text{nm})^{2}$. But if we used the maximum electrostatic positive charge $\sim +5\,(R_{\rm grain}/\text{nm})$, or the intermediate collisional $q_{d}/e \sim -1.4\,(R_{\rm grain}/\text{nm}) (T/10^{4}\,{\rm K})$, or photo-electric charging for HII regions, it gives the same order-of-magnitude in Eq.~\eqref{eq:ratiocomp}.} 
\citep{draine1987collisional, tielens2005physics, draine2010physics}. Here $f_{q} \equiv \sigma_{ed}/\sigma_{d+} \sim \exp{(-\psi)}/(1+\psi)$ is the Coloumb correction factor, accounting repulsion of electrons and focusing for protons assuming mostly negatively dust grains, as detailed by \citet{weingartner1999interstellar}. The collisional charging factor $\psi \equiv Z_{d}q_e^2/k_B T R_{\rm grain} \sim 3.3$ gives $f_{q} \sim 0.008$ \citep{draine1987collisional}.

In Eq.~\eqref{eq:ratiocomp}, the term $|q_{d} \sigma_{e+}/q_{+}\sigma_{ed}|$ can be large at temperatures $T \lesssim 10^{5}$\,K. This is partly because of grain multiple-charging (for larger grains), but mostly because the Spitzer collision rate $\sigma_{e+}\sim 10^{-12}\,{\rm cm^{-2}}\,(T/10^{4}\,{\rm K})^{-2}$ is large \citep{spitzer:conductivity, pinto.galli:2008.momentum.transfer.coefficients.for.weakly.ionized.systems}. Physically, this means electrons are coupled strongly to ions, which -- like neutral collisions in the mostly-neutral limit (\S~\ref{sec:rad.battery.neutral}) -- makes them behave (relative to grains) more like a single electron-ion fluid which is much less mobile and has a much lower effective charge-to-mass ratio. Therefore, this enables easier charge separation, explaining the strong inverse temperature dependence.

 To estimate the electric field in the intermediate temperature regime ($T \lesssim 10^{5}$\,K), we return to the full expression in Eq.~\eqref{eqn:ebat.ion}\footnote{At low dust abundance or intermediate temperatures, the full Eq.~\eqref{eqn:ebat.ion} is needed. At high temperatures $T \gg 10^{5}$\,K (if dust can survive at all), where the $\omega_{e+}$ term becomes negligible in both numerator and denominator of Eq.~\eqref{eqn:ebat.ion}, the predicted battery strength drops rapidly to $\sim 10^{-22}\,{\rm statV\,cm^{-1}} (R_{\rm grain}/\text{nm})^{-1} (Z/Z_{\odot})$}. In this limit, the $\sigma_{e+}$ term dominates the numerator, while the $\sigma_{d+}$ term, dominates the denominator, simplifying the expression for the electric field to: 
\begin{align}
\label{eq:E.ionized}
{\bf E}^{\prime}_{{\rm bat},d} \rightarrow & -\frac{n_{d}q_{d} \omega_{e+}\delta {\bf a}_{d,g}}{\mu_{e} \omega_{d+}} \\ \nonumber &\sim -\frac{n_d q_d \sigma_{e+}}{n_e q_e^2 (1+\psi) \sigma _{d+}}\sqrt{\frac{m_e}{m_+}}\frac{{\bf F}_{\rm rad} \sigma_{d,r}}{c} \\& \nonumber \sim 10^{-20} \, {\rm statV \, cm^{-1}} \, \left( \frac{q_{d}}{q_{d}^0} \right) \left( \frac{ \rm Z}{\rm Z_{\odot}} \right) \\
& \nonumber\quad \times \left( \frac{1 \, {\rm nm}}{R_{\rm grain}} \right) \left( \frac{T}{10^{4} \, {\rm K}} \right)^{-2}  \left( \frac{{\bf F}_{\rm rad}}{{\rm erg \, cm^{-2} \, s^{-1}}} \right).
\end{align}
This solution is identical to that derived for the arbitrary dust velocity in Eq.~\ref{eqn:alpha.limits1} and parallels the mostly neutral limit, with ${\bf J}_{d} \rightarrow n_{d} q_{d} \delta {\bf u}_{d, g} \rightarrow n_{d} q_{d} \delta {\bf a}_{d, g} /\omega_{d+}$, i.e., the dust dynamics are primarily governed by the collision or stopping time, $t_{S} \sim \omega_{d+}^{-1}$.

However when $\omega_{e+}$ dominates in the numerator \textit{and} the denominator, which occurs when $\sigma_{e+}/ \sigma_{d+} > \sqrt{m_{+}/m_{e}}\,(n_{e}/n_{d} Z_{d}^{2})$, we obtain: 
\begin{align}
     &{\bf E}^{\prime}_{{\rm bat},d} \rightarrow -\frac{m_{d} \delta {\bf a}_{d,g}}{q_{d}} \sim -\frac{{\bf F}_{\rm rad} \sigma_{d,r}}{q_{d} c}  \\\nonumber & \sim 2 \times 10^{-15} \,{\rm statV\,cm^{-1}} \left( \frac{R_{\rm grain}}{{1 \rm nm}}\right)^{2}\,\left( \frac{{\bf F}_{\rm rad}}{{\rm erg \, cm^{-2} \, s^{-1}}}\right).
\end{align}

In this limit, electrons and protons are strongly coupled and move together, and charge neutrality forces them to drag along the dust grains, thus modifying the drift speed. Achieving this regime is challenging, however, as it requires extremely low temperatures, specifically $T \lesssim 1 \, {\rm K} \, Z_d  {\rm (Z/Z_{\odot})}^{1/2}  (R_{\rm grain}/{\rm nm})^{1/2}$.

As in \S~\ref{sec:rad.battery.neutral}, we can use the expression in Eq.~\eqref{eq:E.ionized} to obtain a dimensional estimate of the magnetic field growth rate around a star or AGN in this limit, 
\begin{align}
\label{eq:seeding.ion}
\frac{\partial {\bf B}}{\partial t} &= -c\nabla \times {\bf E}'_{\text{bat},d} \\ 
&\sim \frac{0.7\, {\rm nG}}{\rm yr} 
    \left(\frac{L}{L_{\odot}}\right) 
    \left(\frac{\rm AU}{R} \right)^{3} 
    \left(\frac{R}{\ell}\right) \notag \\
& \nonumber \quad \quad \times 
    \left( \frac{1 \, {\rm nm}}{R_{\rm grain}}\right) 
    \left( \frac{T}{10^{4} \, {\rm K}} \right)^{-2} 
    \left( \frac{ \rm Z}{\rm Z_{\odot}} \right) \\ 
&\nonumber \sim \frac{8 \times 10^{-14}\, {\rm G}}{\rm yr} 
    \left(\frac{L}{10^{12} L_{\odot}}\right) 
    \left(\frac{\rm pc}{R}\right)^{3} 
    \left(\frac{R}{\ell}\right) \notag \\
& \nonumber\quad \quad \times 
    \left( \frac{1 \, {\rm nm}}{R_{\rm grain}}\right) 
    \left( \frac{T}{10^{4} \, {\rm K}} \right)^{-2} 
    \left( \frac{\rm Z}{\rm Z_{\odot}} \right)
\end{align}

As a result, while the strength of the dust battery in the well-ionized limit does depend on the dust-to-gas ratio and grain properties (unlike the neutral limit), we anticipate it could be the dominant battery effect at low temperatures $T \ll 10^{5}$\,K down to metallicities below those of the most metal-poor stars known, $Z \sim 10^{-5}\,Z_{\odot}$ \citep{beers2005discovery, norris2007he, yong2012most, keller2014single, hansen2014exploring, frebel2015sd, bonifacio2015topos, caffau2016topos}.

Note that we can also compare ${\bf E}_{{\rm bat,\,d}}^{\prime}$ to the maximum possible Biermann battery strength. The usual Biermann expression is obtained by replacing $\delta {\bf a}_{e,g}$ appropriately in Eq.~\eqref{eq:eom} with the ${\bf G}_{e}$ term in Eq.~\eqref{eqn:drift.species}, assuming a thermal equilibrium distribution function, that the electrons carry a vanishingly small fraction of the inertia, and that the pressure gradient length scales also vary on scales $\sim \ell$.  To quantify the misalignment required in the Biermann mechanism between the electron temperature and density gradients—necessary for producing a non-vanishing $\partial_{t}{\bf B}$-we define $|\sin{\theta}| \equiv |\nabla n_{e} \times \nabla T_{e}| / |\nabla n_{e}| \, |\nabla T_{e}|$. With these limits, the ratio of the Biermann strength to the radiative electron battery, scaled to the same values of incident flux as in e.g.\ Eq.~\eqref{eq:seeding}, is $|{\bf E}_{\rm bat,\,Bier}^{\prime}|/|{\bf E}_{{\rm bat},e}^{\prime}| \sim 10^{-6}\,|\sin{\theta}|\,(R/{\rm pc})\,(R/\ell)\,(L/10^{12}\,L_{\odot})\,(T/{\rm 10}\,K)$. This becomes completely negligible for $\gtrsim$\,pc-scale coherent ${\bf B}$. Even this ignores various corrections that should appear to suppress the Biermann term (and $|\sin{\theta}|$) in the strongly-collisional limit, but still shows that it is at most comparable to the radiation electron battery, and therefore many orders of magnitude weaker than the dust battery unless we consider both very small-scale modes $\lesssim$\,AU in much hotter gas $T \gg 10^{4}$\,K.

 \subsubsection{Favorable Conditions and Minimum Dust Masses}
\label{sec:favorable}

\begin{figure*}
    \centering
\includegraphics[width=1\linewidth]{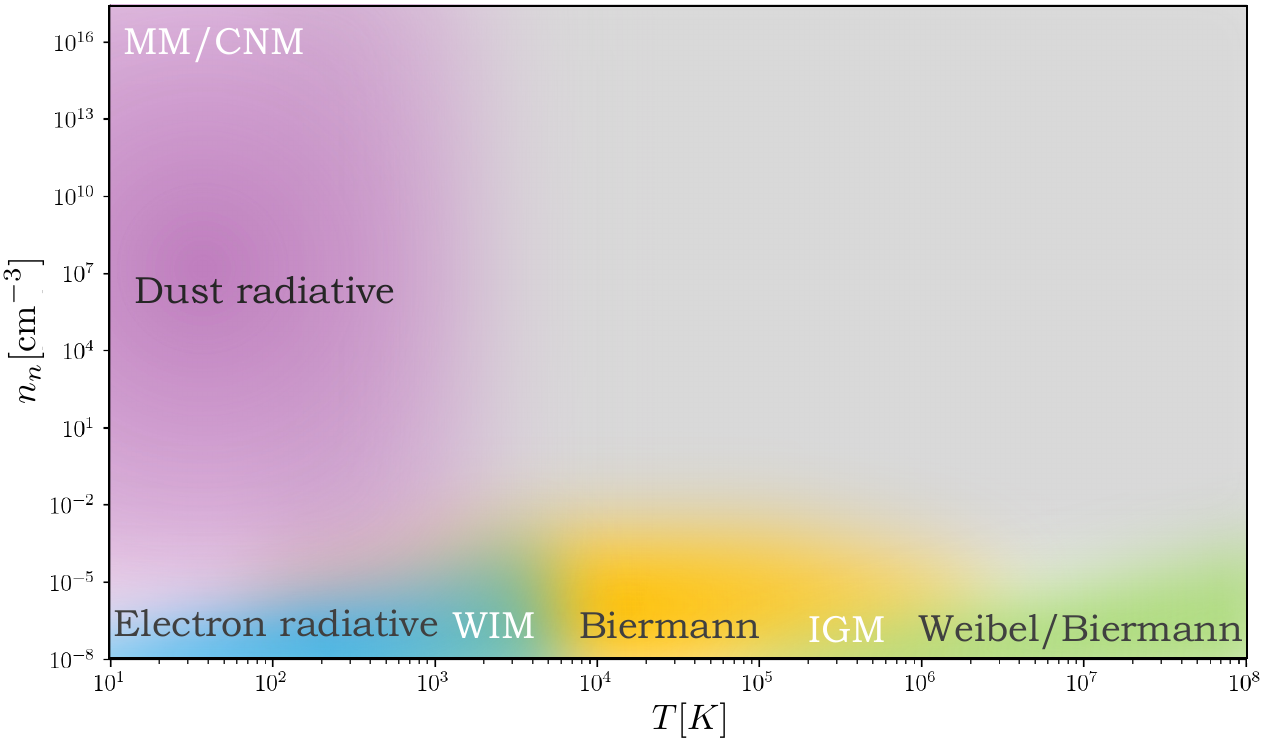}
\caption{Approximate illustration of the dominant battery mechanisms across temperature and density regimes. Shaded regions represent typical environments where each mechanism is most effective. For the electron and dust radiative battteries, we consider a radiation field corresponding to one solar luminosity at 1 AU. We find that for metallicities $\rm Z \sim 10^{-5} Z_\odot$, the dust battery dominates in star-forming regions and cool molecular gas (cold neutral medium (CNM) and molecular medium (MM)) at \(T \lesssim 10^3 \, \text{K}\) and \(n_n \gtrsim 10^4 \, \text{cm}^{-3}\). At lower densities, roughly corresponding to the warm ionized medium (WIM), the radiative electron battery dominates due to higher ionization fractions. In fully ionized plasmas at higher temperatures and lower densities ($\sim 1-10^3 \, \text{cm}^{-3}$), typical of the outer regions of massive galaxies, shocks and the intergalactic medium (IGM), Biermann and Weibel instabilities prevail, with Weibel generally dominating in higher-energy, higher-density environments. The top right corner represents high temperature and high density environments, characteristic of stellar photospheres, alongside more extreme conditions akin to those found inside neutron stars, where plasma behavior differs significantly. However, these conditions are expected to be extremely rare.}
\label{fig:comp}
\end{figure*}
\label{sec:neutrality}

In practice, ISM chemistry and grain charging is quite complex, and large multi-species chemical networks like those in \citet{glover2010modelling,wurster2016nicil,xu:2019.ppd.chemical.networks.for.nonideal.mhd} are often used to calculate the abundances of various species. These also account explicitly for a size spectrum of grains (with multiple positive or negative grain charge values), ionization by cosmic rays and radioactive decay (collisions can also be important at higher temperatures, and photo-ionization in the strong radiation environments of interest). The values and scalings in this paper can be calculated for such arbitrarily long list of species using the approach in Appendix~\ref{sec:general.solution.appendix}, if desired. If we do so for the representative equations in the text, taking the values of different abundances from \citet{wurster2016nicil}, we obtain similar quantitative results (at the order-of-magnitude level) in all cases. The reason is simply that despite the complexity of these networks, charge balance and currents are usually given to leading order by a combination of free electrons, a dominant ion species, and the dust smallest grains. 

We can also motivate some of our implicit assumptions from these chemical network codes. For example, in \S~\ref{sec:rad.battery.neutral}, we derived expressions generically valid in the mostly-neutral regime (applicable when $\omega_{en} > \omega_{e+}$), but often further simplified these by assuming free electrons were depleted onto grains, where the grain properties factor out entirely. The networks above can predict when this occurs, but it clearly requires a sufficient number of grains to ``hold'' the charge (it cannot be true if there is no dust). Taking a simple ionization rate $\zeta \sim \zeta_{-17}= 10^{-17}\,{\rm s^{-1}}$ (scaled to the local Solar neighborhood value), dust-to-gas ratio scaled with metallicity $Z$, and toy three-species ionization-recombination-dust attachment balance model from \citet{keith2014accretion}, electron depletion on dust grains requires $Z\gtrsim 10^{-5}\,Z_{\odot} (R_{\rm grain}/\text{nm})^{2} \zeta_{-17}^{1/2} (T/10^{4}{\rm K})^{-1/4}\,(n_{n} / 10^{10}\,{\rm cm^{-3}})^{-1/2}$, similar to the values obtained in the more detailed chemical networks above. But even without complete electron depletion, we showed in Eq.~\eqref{eqn:ebat.neutral} that the dust battery is stronger than the electron battery for $|n_{d} q_{d} | \gtrsim 10^{-8} |n_{e} q_{e}|$, which the same calculation above shows is easily satisfied even at $Z \sim 10^{-5} Z_{\odot}$ for any plausible ionization rate and grain size, at any gas density where the system could plausibly be mostly-neutral (e.g.\ $n\gtrsim 0.1\,{\rm cm^{-3}}$). And we noted that in most of the well-ionized regime, the dust battery could be important at $Z \gtrsim 10^{-5}\,Z_{\odot}$ as long as $T \lesssim 10^{4}\,$K. These conditions are broadly expected to be realized in star-forming, ISM gas almost immediately after the very first generation of Pop III star formation (i.e.\ before second-generation star formation; see \citealt{wise2011birth, chiaki2019seeding}). Further, it is widely-believed that dust is required for the formation of observed hyper-metal-poor stars \citep{klessen2012formation, nozawa2012can, ji2014chemical, hopkins2017formation}.

\begin{figure}[h!]
\includegraphics[width=\columnwidth]{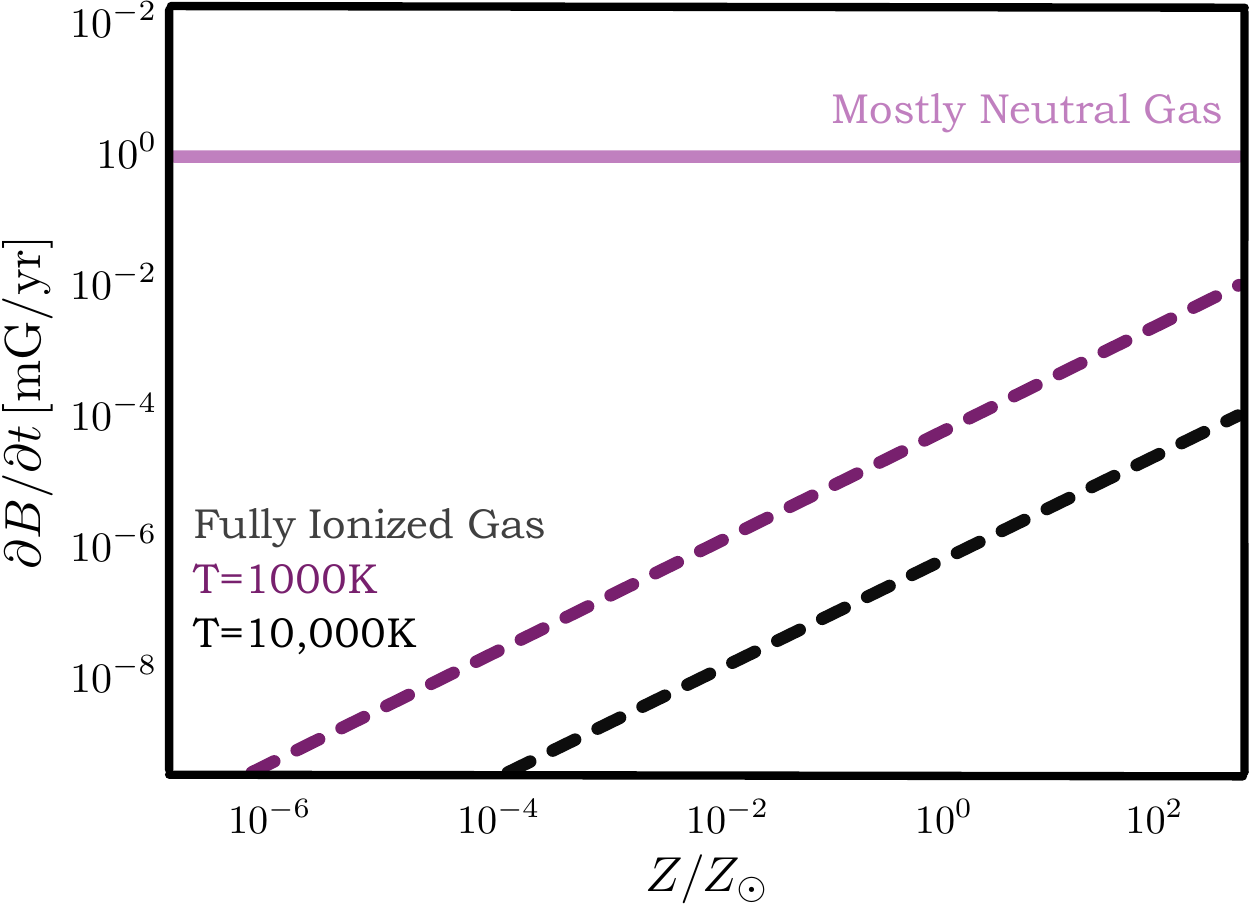}
    \caption{Seeding rates for the dust battery as a function of metallicity for both neutral and ionized environments. In the ionized case, we present seeding rates for two temperatures,  T = 1000K  and  T = 10000K, to illustrate the temperature dependence of the dust battery mechanism. These rates are calculated for a radiation field corresponding to solar luminosity at a 1 AU radius, assuming 1 nm grains and AU-scale gradients.}
    \label{fig:metal}
\end{figure}

We provide a summary of the dominant battery mechanisms across various temperature and density regimes, as shown in Figure \ref{fig:comp}, comparing the dust battery with the radiative electron battery, Biermann battery, and Weibel instabilities. We find that the figure is largely insensitive to variations in the radiation flux  $F_{\text{rad}}$ , as the relative strengths of the dust and electron batteries remain independent of its absolute value. In high-temperature and strongly ionized environments, where dust sublimates, only the Biermann and Weibel instabilities are relevant for comparison with the electron battery. Since these instabilities dominate in the ionized regime under most conditions, changes in radiation flux do not significantly affect the results. Additionally, the figure is mostly independent of metallicity, provided that the metallicity exceeds  $10^{-5}$  in the ionized case, and is largely unaffected by metallicity in the neutral case. For clarity, however, we illustrate the seeding rate as a function of metallicity for both the ionized and neutral cases in Figure \ref{fig:metal}.

 For a given ionization rate, lower neutral densities result in higher ionization fractions, leading to a greater abundance of free electrons relative to charged dust. But at low temperatures, electron-ion coupling ensures the dust battery remains effective (\S~\ref{sec:solutions.ionized}. So there is a narrow ``wedge'' at high ionization (low densities) and intermediate temperatures where the radiative electron battery could dominate. At higher temperatures and lower densities ($n \ll 0.01-1\,{\rm cm^{-3}}$ and $T \gg 10^{4}\,$K), dust abundances are low (especially because this often refers to the circum-and-intergalactic medium, where the abundance of dust even at low redshifts may be low, though see \citealt{holwerda2009extended, menard2010measuring,menard2012cosmic, peek2015dust, wendt2021muse}), the plasma is well ionized, and it is increasingly weakly collisional (e.g.\ Spitzer collision rates drop rapidly). In this regime, we expect the Biermann and Weibel mechanisms to dominate (see references in \S~\ref{sec:intro}), with the latter more important (on macroscopic astrophysical scales) in even more weakly-collisional (higher-$T$, lower-$n$ environments).

\subsection{A sub-grid model for magnetic field seeding in cosmological simulations}
\label{sec:subgrid}

As argued in \S~\ref{sec:general.equations}, the battery scenario here can be studied in future MHD-PIC simulations that resolve ISM dynamics. This includes turbulence, and the associated gradients in dust distributions and radiation fields, which would in turn inform the non-linear efficacy of the dust battery and its saturation amplitudes and structure on different scales. This is especially important for understanding the efficiency of small-scale fields growing into larger-scale magnetic fields relevant for cosmological and galactic (let alone circum/inter-galactic) magnetization. 

However, in the absence of such studies, it is worth considering a (speculative) sub-grid model for ``seeding'' magnetic fields in cosmological simulations via the radiation-dust battery, given that in the typical cosmological simulation the temporal, spatial, and density scales of interest here (e.g.\ scales surrounding individual stars, dusty ``torii'' around AGN, multi-phase structure and resolved turbulence in the ISM) are largely unresolved. We are heuristically motivated by analogy to ad-hoc ``stellar seed'' (or ``supernova seed'' or ``AGN seed'') models, where in a cosmological simulation seed fields are simply injected in some set of macroscopic resolution elements surrounding ``star'' or ``black hole'' particles \citep[see e.g.][]{beck2013strong, butsky2017ab, garaldi2021magnetogenesis, ntormousi2022closer}. 

One option is to simply add a source term ${\bf E}^{\prime}_{{\rm bat},d}$ to $\partial_{t}{\bf B}$, i.e.\ implement Eqs.~\eqref{eq:seeding} \&\ \eqref{eq:seeding.ion}, in terms of some simply on-the-fly estimate of the incident flux ${\bf F}$ in a cell, and an assumed minimum resolvable mode scale $\ell \sim \Delta x$, above some minimum threshold metallicity per \S~\ref{sec:favorable}. Because the growth is so rapid, however, this would naively give unphysically large ${\bf B}$ integrated to resolved timescales (often $\gtrsim $\,Myr). This suggests injecting some saturation estimate of ${\bf B}$ on scales $\Delta x$. If we assume the gas of interest is highly neutral, this can come from Eq.~\eqref{eqn:bsat}, or $|{\bf B}| \sim 5\,{\rm mG}\,(L/10^{10}\,L_{\odot})\,({\rm kpc}/\Delta x)\,\langle n_{e}/10^{-15}n_{n}\rangle (T/10^{4}\,{\rm K})^{-1/2}$, but outside of the most poorly-ionized environments this will again give very large values. But assuming the turbulent resistivity expressions therein gives a more reasonable scaling, the implied saturation $|{\bf B}| \sim 10^{-9}\,{\rm G}\,(F_{\rm rad}/{\rm erg\,s^{-1}\,cm^{-2}})\,({\rm km\,s^{-1}}/v_{\rm turb}[\Delta x])$ on scales $\Delta x$. So this would effectively set a ``floor'' to $|{\bf B}|$ in regions of finite $F_{\rm rad}$ and $\rho_{d}$ (and $F_{\rm rad}$ can be estimated from whatever local sources, i.e.\ star particles or black holes, are in the simulation). 

A broadly similar seed strength is obtained if we integrate over a population of stars with an observed \citet{kroupa2001} initial mass function, assuming they are a zero-age population (each star following the luminosity-mass relation) for some effective mean lifetime $\tau \sim 10\,$Myr, each rapidly amplifying the fields on some smaller micro scales $\ell\sim {\rm AU}$ around the source (where $F_{\rm rad}$ is locally large). This then isotropically expands (along with stellar winds, supernovae, and other ``feedback'') to fill some numerical injection radius $r_{\rm inj}\sim \Delta x$. This gives:

\begin{align}
    \ {{B}_{\rm \star, inj}}&=  \tau \left( \frac{R}{r_{\text{inj}}} \right)^2 { \int \xi(m) \, B_{\star} (m) \, dm } \\
    \nonumber &\sim  10\,{\rm nG} \left (\frac{\tau}{\rm 10\,Myr} \right )\left( \frac{\Delta m_{\ast}}{1000 M_\odot}\right) \left(\frac{\rm AU}{\ell} \right) \left( \frac{ \text{1 kpc}}{r_{\text{inj}}}\right)^{2}\\
    \nonumber &\sim 40\,{\rm nG} \left (\frac{\tau}{\rm 10\,Myr} \right ) \left( \frac{L^{\rm ZAMS}}{10^6 L_\odot}\right)\left( \frac{\rm AU}{\ell} \right) \left( \frac{1 \, \text{kpc}}{r_{\text{inj}}}\right)^{2},
\end{align}
in terms of either the total stellar mass formed $\Delta m$ or the zero-age main sequence luminosity $L^{\rm ZAMS}$ of the population. The latter allows one to easily rescale this to AGN or other sources, assuming some duration.


\section{Comparison to other battery mechanisms}
\label{sec:comp}

The dust radiative battery can generate relatively strong magnetic fields, \( B \sim \) mG, around stars and active galactic nuclei (AGN) over scales as large as $\gg$\,AU-pc. In this section, we compare some aspects of this mechanism with those of traditional, more widely-studied battery mechanisms.

The Biermann battery operates in environments characterized by misaligned gradients of temperature and electron number density, such as ionization fronts and oblique shocks. It can produce magnetic fields with strengths around \( B \sim 10^{-21} - 10^{-19} \ \rm G \) over parsec to kiloparsec scales \citep{kulsrud1997protogalactic}. However, it remains uncertain whether these small seed fields can be amplified to match the observed present-day magnetic field strengths \citep{kulsrud_zweibel}. Additionally, as noted in \S~\ref{sec:favorable} \&\ \ref{sec:solutions.ionized}, the Biermann battery predominantly functions in ionized, warm/hot, and relatively low-density regions, which might not have been abundant in the early Universe. 
It is also prone to self-quenching \citep{ryu1998cosmic, gnedin2000generation}.

Kinetic instabilities, such as the Weibel instability, offer another pathway for magnetic field generation and amplification by exploiting anisotropies in the velocity distribution of particles \citep{quataert2015linear, schoeffler2016generation, zhou2023magnetogenesis, sironi2023generation}. Although this mechanism can generate strong fields (nG), they are correlated on small scales, on the order of the ion-skin depth (\( \sim 10^{-8} \) pc). As a result, these fields average out to weak seed fields over the larger scales relevant to galactic dynamos \citep{kato2008nonrelativistic, chang2008long}. Additionally, kinetic instabilities are effective in extreme, collisionless environments (the hottest, lowest-density astrophysical plasmas), corresponding to the virialized, shocked gas around the most massive halos at the lowest redshifts \citep{califano1997spatial, medvedev1999generation, silva2003interpenetrating}; however, their efficacy diminishes in cooler, less energetic environments, potentially limiting their applicability in the early Universe \citep{califano2006three, medvedev2006cluster, silva2021weibel}. Indeed, gas in a typical halo (defined as halos of mass $M_{\rm halo}^{\ast}(z)$, the virial mass corresponding to a $+1\sigma$ cosmological fluctuation at redshift $z$) will not enter the ``Weibel'' parameter space in Fig.~\ref{fig:comp} until redshifts $z\lesssim 0.5$, and even the most massive collapsing regions cannot reach this parameter space until $z\lesssim 3$ \citep{birnboim2003virial, kerevs2005galaxies}.\footnote{One can also see this by noting that the electron-ion mean free path in virial-shocked gas scales as $\lambda_{e+} \propto T^{2}/n_{e} \propto (M_{\rm halo}^{\ast}[z])^{4/3}/(1+z)$, which declines as $\sim (1+z)^{-7}$ until redshift $\sim 3$ and then exponentially,  decreasing by a factor $\gtrsim 10^{15}$ from $z=0$ to $z\sim 10$.}

As shown above, radiation batteries on electrons, driven by Thompson scattering from starlight (or, as originally imagined, the CMB; \citealt{harrison1973magnetic, ando2010generation}) are almost always much weaker, by many order of magnitude than dust radiation batteries. As a result, the radiative-electron battery typically produces quite weak fields (e.g.\ $|{\bf B}| \sim 10^{-23}-10^{-19}\,{\rm G}$; \citealt{matarrese2005large, gopal2005generation}), and even this only in well-ionized, relatively warm gas (\S~\ref{sec:favorable}). It has been suggested that the efficacy of the electron radiation battery could be boosted by ionization effects, specifically accounting for the large opacity of neutral gas suddenly exposed to ionizing radiation \citep{durrive2015intergalactic, durrive2017mean}. But this process is restricted to extremely thin skin depths in ionization layers, limiting its effectiveness \citep{garaldi2021magnetogenesis}. Further, this mechanism is theoretically  ambiguous even within those environments owing to the fact that it does not depend on continuous acceleration term like ${\bf G}$ or ${\bf a}$ in Eq.~\eqref{eqn:drift.species}, but on injection of ``new'' free electrons (a number density source term).

All the aforementioned mechanisms function within restricted environments that are scarce in the early Universe. Consequently, it remains unclear whether they can account for the observed ubiquity of magnetic fields. In comparison, the dust radiative battery has several interesting properties. It can operate in a range of astrophysical environments, including regions with high densities, strong collisional coupling/fluid limits, predominantly neutral gas, significant dust content, and low temperatures, where other mechanisms might be ineffective. We argue that the dust radiative battery could generate seed fields that are several orders of magnitude stronger than those produced by other mechanisms in these regimes, and does not saturate due to Ohmic or turbulent resistivity until large magnetic field strengths (potentially $\gg{\rm \mu G}$) are reached on macroscopic scales. It can also in principle act coherently on much longer length scales than the mechanisms above, depending on the properties of the radiation source. 

One intuitive (albeit over-simplified) way to think of the efficacy of the dust battery, compared to the other batteries discussed above, is in terms of the mean-free-paths of the different particle types. Mechanisms like the Biermann, Weibel, and radiative electron battery are limited in their ability to generate appreciable charge separation on scales much larger than the electron deflection length/mean free path, which is often astrophysically very small. But (in addition to having a large cross section to radiation for acceleration) dust grains have vastly larger deflection/stopping/collisional mean free paths and so it is comparatively easy to induce large dust drift velocities, and hence charge separation.

\section{Conclusions}
\label{sec:conc}

In this work, we have introduced and analyzed a novel mechanism for generating small seed magnetic fields in the early universe through a ``dust battery'' process. This mechanism operates by radiatively accelerating charged grains, leading to charge separation and the subsequent generation of a magnetic field. This work highlights several key findings:

\begin{itemize}
    \item {\bf Efficiency in Diverse Conditions}: The dust battery can operate effectively in predominantly neutral environments near bright sources such as stars, supernovae, and active galactic nuclei, similar to conditions found around second-generation stars in the early universe. This mechanism can generate magnetic fields in the range of nG to $\rm \mu$ G on scales from AU to kpc over timescales of years to Myr.

\item {\bf{Robustness Against Dissipation}}: The dust battery mechanism appears resilient against dissipation effects, such as Ohmic dissipation. Simple estimates suggest that saturation could occur at field strengths up to mG for a radiation field corresponding to one solar luminosity at 1 AU, with T = 10 K in a predominantly neutral gas. However, as shown in Figure \ref{fig:metal}, the seeding rate decreases significantly when the gas is ionized, leading to lower saturation field strengths. Moreover, since saturation occurs in the nonlinear regime, which is yet to be explored through simulations, this estimate should be interpreted with caution. Nonetheless, this relatively large saturation amplitude suggests that   Nonetheless, the relatively large saturation amplitude suggests that the dust battery mechanism can likely sustain the growth of small-scale magnetic field fluctuations, making it a compelling candidate for the origin of seed magnetic fields.

\item {\bf{Comparison to other mechanisms}}: Compared to mechanisms like the Biermann battery, Weibel instability, and radiation-driven electron batteries, the dust battery has distinct properties. Those mechanisms typically require specific conditions—such as shocks, high ionization fractions, and high-energy/temperature and low-density (weakly-collisional) environments—that were likely confined to rare regions with small filling factors in the early Universe, and may not be able to generate coherent seed fields on scales larger than astrophysical ``microscales.'' In contrast, the dust battery is most effective in predominantly neutral gas, even at low metallicities (\( Z \sim 10^{-5} Z_\odot \)), and can likely seed magnetic fields efficiently in the vicinities of stars or AGN with coherence lengths in principle extending up to $\sim\,$pc scales. This is well into the scales where more traditional dynamo mechanisms and turbulence could further amplify said fields. To summarize the distinctions between various battery mechanisms:  

 \begin{enumerate}
    \item \textbf{Dust Battery}: Can plausibly generate $\sim \rm mG$ fields, with spatial scales determined by dust density fluctuations and radiation field gradients, without a strict upper limit. Most efficient in neutral gas but also operates in ionized regions.

    \item \textbf{Biermann Battery}: Generates $B \sim 10^{-21}-10^{-19} \,\text{G}$ on parsec to kiloparsec scales, efficient in ionization fronts and oblique shocks.  \\
    \item \textbf{Weibel Instability}: Produces $B \sim \text{nG}$ on ion skin depth scales ($\sim 10^{-8} \,\text{pc}$), relevant in collisionless shocks and anisotropic plasmas.

    \item \textbf{Electron Battery} – Generates $B \sim 10^{-23}-10^{-19} \,\text{G}$ in well-ionized environments on parsec to kiloparsec scales.

\end{enumerate}
\end{itemize}

Overall, the dust battery mechanism appears to be an interesting candidate for generating seed magnetic fields in the early Universe, at least in some environments. And we stress that nothing here argues that other battery mechanisms do not occur -- but given the very different conditions under which they operate efficiently, it may well be that multiple battery mechanisms operate at different times and places. Further investigation is needed to fully understand the implications and integration with other processes of the dust battery in cosmic magnetic field evolution. We present the equations needed to implement a magnetohydrodynamic-particle-in-cell method (MHD-PIC) and sub-grid model for cosmological simulations, allowing for the study of the efficacy, non-linear behavior/saturation, and evolution of this mechanism on both the salient plasma scales as well as macroscopic cosmological scales (where the sub-grid models for the latter can be informed by the former). Future work should focus on exploring these aspects in greater detail to assess the role of the dust battery mechanism more comprehensively.

\vspace{1em}
 \noindent Support for for NS and PFH was provided by NSF Research Grants 1911233, 20009234, 2108318, NSF CAREER grant 1455342, NASA grants 80NSSC18K0562, HST-AR-15800. JS acknowledges the support of the Royal Society Te Ap\=arangi, through Marsden-Fund grant  MFP-UOO2221 and Rutherford Discovery Fellowship  RDF-U001804.

\bibliography{sample631}{}
\bibliographystyle{aasjournal}


\begin{appendix}

\section{General solution for an  $N$-Species Fluid}
\label{sec:general.solution.appendix}



In the main text, we derived the governing equations for a four-species fluid in the limit of $\mathbf{B} = 0$ (or for the velocity components parallel to $\mathbf{B}$). Here, we generalize this derivation to account for all velocity components in a fluid with an arbitrary number of species, $N$. We start from Eq.~\eqref{eqn:eom.all} in the main text, which describes the momentum exchange between species $j$ and the surrounding fluid:
 \begin{align}
 \label{eqn:eom.appendix}
     - \delta {\bf{a}}_{j, g}   =  \frac{q_{j}}{m_{j}} {\bf E}^{\prime}+  \Omega_{j} \delta {\bf u}_{j,g} \times \bhat  + \sum_{i} \omega_{ji} ( \delta {\bf u}_{i,g} - \delta {\bf u}_{j,g} ) .
\end{align}

As emphasized in the main text, the terminal velocity approximation does not apply to neutral particles. Therefore, we use the definitions for the bulk velocity $\mathbf{U}_g$ and the drift velocity dispersion $\delta \mathbf{u}_{j, g}$, combined with the quasi-charge neutrality condition. This leads to the following constraint:
\begin{align}
    \sum_{j} \rho_{j} \delta {\bf u}_{j,g} =  \sum_{j} n_{j} q_j\delta {\bf u}_{j,g}=\boldsymbol{0}.
\end{align}
These constraints, together with Eq.~\eqref{eqn:eom.appendix}, close the system of equations for the drift velocities $\delta \mathbf{u}_{j, g}$ and the electric field $\mathbf{E}^{\prime}$. When the coefficients $\omega_{ji}$ and $q_j$ are weakly dependent on the drift velocities, this system is linear, enabling us to decompose the solution for $\mathbf{E}^{\prime}$ into two components: $\mathbf{E}^{\prime} = \mathbf{E}^{\prime}_{J} + \mathbf{E}^{\prime}_{\text{bat}}$. Here, $\mathbf{E}^{\prime}{J}$ corresponds to the electric field component driven by the current, while $\mathbf{E}^{\prime}{\text{bat}}$ represents the battery-driven component (with no current, $\mathbf{J} = \mathbf{0}$). 

To solve for the electric field $\mathbf{E}^{\prime}{\text{bat}}$ and the drift velocities $\delta \mathbf{u}{j,g}$ for an arbitrary number of species N, we recast the system into a matrix equation. Specifically, we rewrite the equations of motion as
\begin{align}
    {\bf M} \cdot {\bf V} = {\bf A},
\end{align}
where ${\bf V}$ contains the unknowns, $\delta {\bf u}){j,g}$ and $\mathbf{E}^{\prime}_{\text{bat}},$ and ${\bf A}$ represents the acceleration terms, ${\bf A} = \{{\bf 0}, …, -\delta {\bf a}_{j, g}, \ldots\}$. The matrix ${\bf M}$, which is generally invertible, is a $3(N+1) \times 3(N+1)$ matrix that encapsulates the system’s dynamics, allowing us to solve for ${\bf V} = {\bf M}^{-1} \cdot {\bf A}$.

Let us denote ${\bf B }= \left ( B_x, B_y, B_z\right)$ and ${\bf{E}}_{{\rm bat}, d}= \left ( E^\prime_{{\rm bat}, x}, E^\prime_{{\rm bat}, y},  E^\prime_{{\rm bat}, z}\right)$. The system of equations for each species j can then be expressed as follows:
\begin{align}
\begin{cases}
a_{j,x} = \frac{q_j}{m_j} E'_{\text{bat},x} + \Omega_{j,z} \delta u_{j,y} - \Omega_{j,y} \delta u_{j,z} + \sum_i \omega_{ji} (\delta u_{i,x} - \delta u_{j,x}) \\
a_{j,y} = \frac{q_j}{m_j} E'_{\text{bat},y} - \Omega_{j,z} \delta u_{j,x} + \Omega_{j,x} \delta u_{j,z} + \sum_i \omega_{ji} (\delta u_{i,y} - \delta u_{j,y}) \\
a_{j,z} = \frac{q_j}{m_j} E'_{\text{bat},z} + \Omega_{j,y} \delta u_{j,x} - \Omega_{j,x} \delta u_{j,y} + \sum_i \omega_{ji} (\delta u_{i,z} - \delta u_{j,z}) \\
\sum_j n_j q_j \delta u_{j,x} =  \sum_{j} \rho_{j}  \delta { u}_{j, x} =
\sum_j n_j q_j \delta u_{j,y} =  \sum_{j} \rho_{j}  \delta { u}_{j, y}= 
\sum_j n_j q_j \delta u_{j,z} =  \sum_{j} \rho_{j}  \delta { u}_{j, z}= 0
\end{cases}
\end{align}

We can group the above system into vector form. For each acceleration component, we define:

\begin{align}
   \mathbf{a}_x = \begin{pmatrix} a_{1,x} \\ a_{2,x} \\ \vdots \\ a_{N,x} \end{pmatrix}, \quad
   \mathbf{a}_y = \begin{pmatrix} a_{1,y} \\ a_{2,y} \\ \vdots \\ a_{N,y} \end{pmatrix}, \quad
   \mathbf{a}_z = \begin{pmatrix} a_{1,z} \\ a_{2,z} \\ \vdots \\ a_{N,z} \end{pmatrix},
\end{align}
with corresponding drift velocity vectors:
\begin{align}
\delta \mathbf{u}_x = \begin{pmatrix} \delta u_{1,x} \\ \delta u_{2,x} \\ \vdots \\ \delta u_{N,x} \end{pmatrix}, \quad
   \delta \mathbf{u}_y = \begin{pmatrix} \delta u_{1,y} \\ \delta u_{2,y} \\ \vdots \\ \delta u_{N,y} \end{pmatrix}, \quad
   \delta \mathbf{u}_z = \begin{pmatrix} \delta u_{1,z} \\ \delta u_{2,z} \\ \vdots \\ \delta u_{N,z} \end{pmatrix}.
\end{align}

We now introduce the key matrices relevant to the system:

The matrix representing the charge-to-mass ratios is given by
\begin{align}
   \mathbf{q} = \text{diag}\left(\frac{q_1}{m_1}, \frac{q_2}{m_2}, \ldots, \frac{q_N}{m_N}\right).
\end{align}

The cyclotron frequency matrices are given by:

\begin{align}
   \mathbf{\Omega}_{x} = \text{diag}\left(\Omega_{1, x}, \Omega_{2,y}, \ldots, \Omega_{N, x}\right), \quad \mathbf{\Omega}_{y} = \text{diag}\left(\Omega_{1, y}, \Omega_{2,y}, \ldots, \Omega_{N, y}\right), \quad \mathbf{\Omega}_{z} = \text{diag}\left(\Omega_{1, }, \Omega_{2,z}, \ldots, \Omega_{N, z}\right).
\end{align}

The interaction matrix, representing collisional interaction terms between species $j$ and $i$, is:
\begin{align}
   \boldsymbol{\omega} = \begin{pmatrix}
   \sum_{i \neq 1} \omega_{1i} & \omega_{12} & \cdots & \omega_{1N} \\
   \omega_{21} & \sum_{i \neq 2} \omega_{2i} & \cdots & \omega_{2N} \\
   \vdots & \vdots & \ddots & \vdots \\
   \omega_{N1} & \omega_{N2} & \cdots & \sum_{i \neq N} \omega_{Ni}
   \end{pmatrix}  .
\end{align}

The density and charge density vectors are: 
\begin{align}
{\boldsymbol{\rho}} = \begin{pmatrix} \rho_1 & \rho_2 & \cdots & \rho_N \end{pmatrix}, \quad 
\mathbf{n} = \begin{pmatrix} n_1 q_1 & n_2 q_2 & \cdots & n_N q_N \end{pmatrix}.
\end{align}

Thus, the full matrix equation becomes:

\[
\begin{pmatrix}
\mathbf{a}_x \\
\mathbf{a}_y \\
\mathbf{a}_z \\
\mathbf{0} \\
\mathbf{0} \\
\mathbf{0} \\
\mathbf{0} \\
\mathbf{0} \\
\mathbf{0}
\end{pmatrix} = 
\begin{pmatrix}
\mathbf{q} & \mathbf{0} & \mathbf{0} & \boldsymbol{\omega} & \mathbf{\Omega}_{z} & -\mathbf{\Omega}_{y}\\
\mathbf{0} & \mathbf{q} & \mathbf{0} & \mathbf{-\Omega}_{z} & \boldsymbol{\omega} & \mathbf{\Omega}_{x} \\
\mathbf{0} & \mathbf{0} & \mathbf{q} & \mathbf{\Omega}_{y} & -\mathbf{\Omega}_{x} & \boldsymbol{\omega} \\
\mathbf{0} & \mathbf{0} & \mathbf{0} & \mathbf{n} & \mathbf{0} & \mathbf{0} \\
\mathbf{0} & \mathbf{0} & \mathbf{0} & \mathbf{0} & \mathbf{n} & \mathbf{0} \\
\mathbf{0} & \mathbf{0} & \mathbf{0} & \mathbf{0} & \mathbf{0} & \mathbf{n}\\
\mathbf{0} & \mathbf{0} & \mathbf{0} & \boldsymbol{\rho} & \mathbf{0} & \mathbf{0} \\
\mathbf{0} & \mathbf{0} & \mathbf{0} & \mathbf{0} & \boldsymbol{\rho} & \mathbf{0} \\
\mathbf{0} & \mathbf{0} & \mathbf{0} & \mathbf{0} & \mathbf{0} & \boldsymbol{\rho}
\end{pmatrix}
\begin{pmatrix}
\mathbf{E}'_{\text{bat},x} \\
\mathbf{E}'_{\text{bat},y} \\
\mathbf{E}'_{\text{bat},z} \\
\delta \mathbf{u}_x \\
\delta \mathbf{u}_y \\
\delta \mathbf{u}_z \\

\end{pmatrix}
\]

Note that while this matrix may seem to have more equations than necessary for solving the system, the last three rows are required to determine the velocities of the neutral species, which are not explicitly present in the other terms.

\end{appendix}

\end{document}